\documentclass[useAMS,usenatbib, letterpaper]{mn2e}

\usepackage{graphicx}
\usepackage{txfonts}
\usepackage{epstopdf}
\usepackage[normalem]{ulem}
\usepackage{natbib}
\usepackage{color}
\usepackage{xcolor, cancel}

\title[Evolution of forced shear flows in polytropic atmospheres]{Evolution of forced shear flows in polytropic atmospheres:\\ 
A comparison of forcing methods and energetics}
\author[V. Witzke, L. J. Silvers and B. Favier]{V. Witzke$^{1}$\thanks{E-mail:Veronika.Witzke.1@city.ac.uk (VW); Lara.Silvers.1@city.ac.uk (LJS); Favier@irphe.univ-mrs.fr (BF)}, L. J. Silvers$^{1}$ and B. Favier$^{2}$ \\
$^{1}$Department of Mathematics, City University London,
              Northampton Square, London, EC1V 0HB, UK\\
$^{2}$Aix Marseille Univ, CNRS, Cent Marseille, IRPHE, Marseille, France}
\begin{document}
%
%\date{Accepted 1988 December 15. Received 1988 December 14; in original form 1988 October 11}
%
\pagerange{\pageref{firstpage}--\pageref{lastpage}} \pubyear{2016}
\maketitle
\label{firstpage}
\begin{abstract}
Shear flows are ubiquitous in astrophysical objects including planetary and stellar interiors, where their dynamics can have significant impact on thermo-chemical processes. Investigating the complex dynamics of shear flows requires numerical calculations that provide a long time evolution of the system. To achieve a sufficiently long lifetime in a local numerical model the system has to be forced externally. However, at present, there exist several different forcing methods to sustain large-scale shear flows in local models. In this paper we examine and compare various methods used in the literature in order to resolve their respective applicability and limitations. These techniques are compared during the exponential growth phase of a shear flow instability, such as the Kelvin-Helmholtz (KH) instability, and some are examined during the subsequent non-linear evolution. A linear stability analysis provides reference for the growth rate of the most unstable modes in the system and a detailed analysis of the energetics provides a comprehensive understanding of the energy exchange during the system's evolution. Finally, we discuss the pros and cons of each forcing method and their relation with natural mechanisms generating shear flows.
\end{abstract}
\section{Introduction}
%---------------------------------------------------------------------------------------------------------------------
%different models: global and local (global includes all physics )
%----------------------------------------------------------------------------------------------------------------------
The relative difficulty of observing most astrophysical shear regions, such as those in the Sun, in detail makes it imperative to use analytical and numerical techniques to shed light on the motions present there. Global-scale numerical calculations of stellar interior dynamics is one approach to investigate the mechanisms maintaining differential rotation \citep{2002ApJ...570..865B, 2008ApJ...673..557M}.
However, by using a global approach, such models can not resolve a large range of length-scales in its entirety and have to rely on artificially large transport coefficients or subgrid-scale models. Therefore, numerical investigations using a local approach, where only a small fraction of the object is simulated, can help to provide a more detailed description of the region of interest.\\[5pt]
%
%------------------------------------------------------------------------------------------------------------------
%Different shear flows in astrophysical objects
%------------------------------------------------------------------------------------------------------------------
%
%
%
Previous studies of astrophysical flows looked at an assortment of different velocity profiles, depending on the problem and choice of boundary conditions. In the case of Keplerian motion, which is investigated in the context of accretion discs, usually a linear velocity profile is assumed (e.g. \citealp{1995ApJ...446..741B}; \citealp*{1999ApJ...518..394H}; \citealp{2005A&A...429....1D, Silvers01042008}). This type of shear profile does not require an external force to balance viscous dissipation and instead they incorporate the velocity via a shearing-box approach (\citealp{1965MNRAS.130..125G}; \citealp*{1987MNRAS.228....1N}), where the velocity is instantaneously present. In contrast, some investigations of stellar shear flows have used polynomial functions, as for example in \citet{0004-637X-603-2-785} and \cite*{2003ApJ...599.1449C} while other investigations have utilized trigonometric functions to model the velocity field \citep*[see, for example,][]{2013JFM...717..395H, 2003ApJ...588..630C}. Such velocity profiles have a non-vanishing gradient at the boundaries. In order to minimise the effect of the boundaries on the shear layer a hyperbolic tangent profile can be used \citep[see for example][]{2001MNRAS.320...73B, 2001RSPSA.457.1365H, 0004-637X-686-1-709}.\\
%------------------------------------------------------------------------------------------
%------------------------
%
%
%---------------------------------------------------------------------------------------------------------------------
%Forcing methods, no linear but non-linear calculations
%----------------------------------------------------------------------------------------------------------------------
Hyperbolic tangent profiles are commonly used in classical studies of Kelvin-Helmholtz instability and turbulence. However, most local numerical studies of the turbulence transition in shear flows take the approach of an unforced flow \citep{2000JFM...413....1C, /10.1063/1.870386, 2003JPO....33..694S}, which results in a finite lifetime of an initially unstable background state due to its inevitable viscous decay. However, astrophysical shear flows can be either transient features or be sustained over very long time-scales, where the physical mechanism maintaining the shear flow is usually unknown.   Incorporating a forcing into the numerical model has therefore two roles first to sustain the initial state, for which a linear stability analysis can be carried out, and second to model the unknown physical processes responsible for the resulting flow in an astrophysical system. A variety of classical studies of shear driven turbulence exploit a method where a decoupled background shear flow is present (for example used by  \citealp*{1992JFM...237..499H, 1997JFM...342..231J}; \citealp{2012MNRAS.424..115B}). This method requires a change of variables to incorporate a mean shear profile and does not allow for a back-reaction of the actual flow on the forcing.
%----------------------------------------------------------------------------------------------------------------------
%----------------------------------------------------------------------------------------------------------------------
Whereas investigations of astrophysical shear flows exploit different methods to provide a sustained flow. For example in \citet{0004-637X-586-1-663}, \citet{0004-637X-686-1-709}, and \citet{1538-4357-702-1-L14}  a method to balance viscous dissipation by introducing an external force proportional to the viscous term is utilized. Another option that has been selected is the relaxation method \cite[e.g.][]{refId0m01}, which incorporates an external force proportional to the difference between the actual velocity profile and the target shear profile.
Furthermore, studies focusing on magnetohydrodynamical instabilities used forced shear flows in order to study magnetic buoyancy and its relevance to the formation of sunspots \citep*{0004-637X-603-2-785, 2002MNRAS.329L..73B}. 
To understand the role of a shear flow for magnetic field generation, several investigations on the interaction between a shear flow and initial weak structured magnetic fields have been conducted (\citealp*{0004-637X-686-1-709, 0067-0049-168-2-337}; \citealp{2015JFM...767..199H}). In all cases however, the influence of the forcing method used on the evolution of the system has not been studied.   \\ 
%%-----------------------------------------------------------------------------------------------------------------
%---------------------------------------------------------------------------------------------------------------------
%structure of the paper
%----------------------------------------------------------------------------------------------------------------------
In this paper, a comparative analysis of different forcing methods is performed to understand how different forcings affect the non-linear evolution of the KH instability. 
The governing equations are given in Sec. \ref{sec:model} along with the formulation of the forcing methods and numerical methods used. Our comparison of the different forcing methods is presented in Sec \ref{Sec:comparison_forceing}, where the exponential growth regime is compared in Sec. \ref{sec:results_linear_compare} and the non-linear phase is presented in Sec. \ref{sec:3.2non-linear}. 
%
%%%%%%%%%%%%%%%%%%%%%%%%%%%%%%%%%%%%%%%%%%%%%%%%%%%%%%%%%%%%%%%%%%%%%
%%%%%%%%%%%%%%%%%%%%%%%%%%%%%%%%%%%%%%%%%%%%%%%%%%%%%%%%%%%%%%%%%%%%%%%%%%%%%%%%%%%%%%%%%%%%%%%%%%%%%%%%%%%%%%%%%%%%%%%%%%%%%%%%%%%%%%%%%%%%%%%%%%%%%%%%%%%%%%
%%%%%%%%%%%%%%%%%%%%%%       M O D E L          %%%%%%%%%%%%%%%%%%%%%%%%%%%%%%%%%%%%%%%%%%%%%%%%%%%%%%%%%%%%%%%%%%%                               %%%%%%%%%%%%%%%%%%%%%%%%%%%%%%%%%%%%%%%%%%%%%
%%%%%%%%%%%%%%%%%%%%%%%%%%%%%%%%%%%%%%%%%%%%%%%%%%%%%%%%%%%%%%%%%%%%%%%%%%%%%%%%%%%%%%%%%%%%%%%%%%%%%%%%%%%%%%%%%%%%%%%%%%%%%%%%%%%%%%%%%%%%%%%%%%%%%%%%%%%%%%
%--------------------------------------------------------------------------------------------------------
\section{Three-dimensional Model}
\label{sec:model}
Although a linear stability analysis is a powerful tool to investigate possible instabilities of shear flows in stellar environment as previously studied in \cite*{2015AandAWitzke}, understanding the complex dynamics requires three-dimensional non-linear calculations. In order set up a system that initially remains comparable to a linear stability problem, which has a non-evolving background state, external forcing is needed. The force aims to maintain the initial conditions corresponding to the equilibrium state as long as non-linear effects are negligible. Our purpose is to investigate whether different forcing methods provide a temporally evolving system, which shows the predicted linear evolution, and in what respect the non-linear evolution depend on the method used.
\subsection{Governing equations, boundary conditions, and background state}
\begin{figure}
\includegraphics[width=82mm]{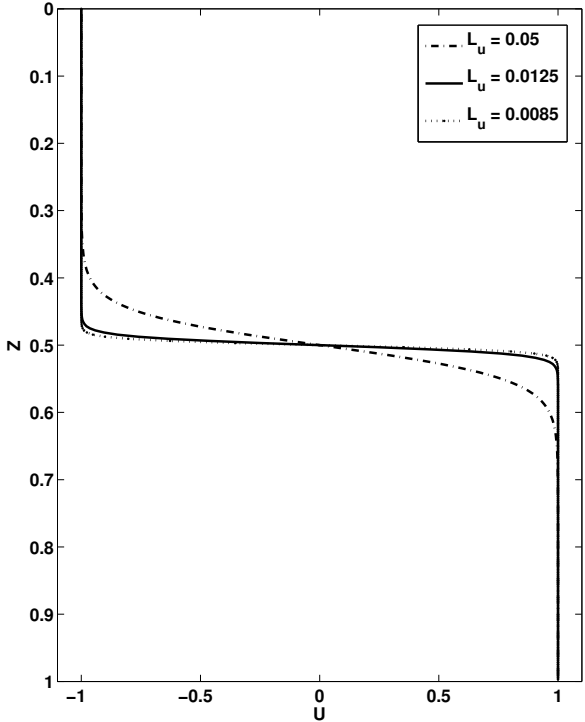}
\caption{Plots of initial shear flow profiles. Shear flow profiles with different $L_u$ parameter, but with the same amplitude $U_0 = 1$ are plotted. The z-axis corresponds to the vertical dimension.}
\label{fig:figure01} 
\end{figure}

We consider a three-dimensional domain of depth $d$, bounded by two horizontal planes located at $z=0$ and $z= 1$, and periodic in both horizontal directions. The fluid is assumed to be an ideal monatomic gas with the adiabatic index $\gamma= c_p/c_v = 5/3$ and constant dynamic viscosity $\mu$, constant thermal conductivity $\kappa$, constant heat capacities $c_p$ at constant pressure, and $c_v$ at constant volume. 
%-----------------------------------------------------------------------------------------------------------------------
The set of dimensionless differential equations we consider is:
\begin{eqnarray}
\label{eq:NSEquation01}
\frac{\partial \rho}{\partial t} & = & - \mathbf{\nabla}\mathbf{\cdot}\left(\rho \mathbf{u} \right) \, \\
\label{eq:NSEquation002}
\frac{\partial(\rho \mathbf{u})}{\partial t} & = & \sigma C_k \left( \nabla^2\mathbf{u}\, +\,\frac{1}{3}\mathbf{\nabla}(\mathbf{\nabla}\mathbf{\cdot}\mathbf{u})\right) -\mathbf{\nabla} \mathbf{\cdot} \left(\rho \mathbf{u u} \right)\, \nonumber \\
 &   & -\,\mathbf{\nabla}p\, + \, \theta(m+1) \rho\, \hat{\mathbf{z}} \, + \, \mathbf{F}\, \\
\label{eq:NSEquation02}
\frac{\partial T}{\partial t}  & = & \frac{C_k \sigma (\gamma -1)}{2\rho}|\mathbf{\tau}|^{2}\,+\,\frac{\gamma C_k}{\rho} \nabla^2 T \nonumber \\
 &  & - \mathbf{\nabla} \mathbf{\cdot}\left(T \, \mathbf{u}\right)\,-\,(\gamma -2)T \mathbf{\nabla} \mathbf{\cdot} \, \mathbf{u}
\label{eq:NSEquations}
\end{eqnarray}
where $\rho$ is the density, $\mathbf{u}$ the velocity field, $T$ the temperature, $\theta$ denotes the uniform temperature gradient across the layer, and $p$ is the pressure.
%---------------------------------------------------------------------------------------------------------------------
In the dimensionless equations above, all lengths are given in units of the domain depth $d$. The temperature and density are recast in units of $T_{t}$ and $\rho_{t}$, the temperature and density at the top of the layer, and we take the sound-crossing time, which is given by $\tilde{t} = d/[ (c_p-c_v)T_{t} ]^{1/2}$, as the reference time. 
There are two dimensionless numbers in the set of equations above: the Prandtl number, $\sigma=\mu c_p/\kappa $, which is the ratio of viscosity to thermal conductivity and the thermal dissipation parameter $C_k= \kappa \tau/(\rho_t c_p d²)$. The strain rate tensor in equation (\ref{eq:NSEquation02}) has the form 
 \begin{equation}
 \tau_{ij}=\frac{\partial u_{j}}{\partial x_i} + \frac{\partial u_i}{\partial x_j} - \delta_{ij} \frac{2}{3}  \frac{\partial u_k}{\partial x_k} .
 \end{equation} 
A force term $\mathbf{F}$ in equation (\ref{eq:NSEquation002}) aims to model external forces 
resulting from large-scale global effects (such as Reynold stresses associated with thermal convection in global-scale calculations for example) not included in our local approach.
This force can sustain a shear flow when needed and is set to zero otherwise. The different forcings that we will consider are described in Section \ref{sec:Forcing_methos}.\\
%----------------------------------------------------------------------------------------------------------
For the basic state a polytropic relation between pressure and density is taken. 
Due to the Schwarzschild criterion the fluid is stable against convection if the inequality $m > 1/(\gamma -1) = 1.5$ holds. In this paper the polytropic index $m$ is always chosen such that the atmosphere is stably stratified. 
The boundary conditions at the top and the bottom
of the domain are impermeable and stress-free velocity and fixed
temperature: 
\begin{eqnarray}
\label{eq:boundary01}
{u}_z = \frac{\partial {u_x}}{\partial z} = \frac{\partial u_y}{\partial z}  = 0  \qquad \textrm{at} \qquad  z = 0   \quad \textrm{and} \quad  z = 1, \\
T=1 \quad \textrm{at} \quad z=0 \quad \textrm{and} \quad  T=1+\theta  \quad \textrm{at} \quad z=1.
\end{eqnarray}
The dimensionless initial temperature and density profiles are of the form:
\begin{equation}
\label{eq:equi_temp}
T(z)= \left(1 + \theta z \right)
\end{equation}
\begin{equation}
\label{eq:equi_dens}
\rho (z) =  \left( 1 + \theta z \right)^{m}.
\end{equation}
This basic state corresponds to an equilibrium state if the fluid is at rest.\\
In order to obtain a shear driven turbulent regime it is necessary to start with an unstable  velocity profile. A hyperbolic tangent profile is assumed in order to model a localised shear layer in the middle or our domain that will minimize the effects of the boundaries. Therefore, we assume that an external force sustains the following initial background velocity profile 
\begin{equation}
\label{eq:target_shear}
\mathbf{U_0} = ( u_0(z), 0, 0 )^T = U_0 \tanh\left(\frac{z-0.5}{L_u}\right) \hat{\mathbf{e}}_x
 \end{equation}
with a shear amplitude $U_0$ and a scaling factor $1/L_u$ that controls the width of the shear profile. The boundary conditions introduced in equation (\ref{eq:boundary01}) restrict the shear profile to values of $L_u$ which will result in a low enough value of the \textit{z}-derivative at the boundaries. Using this velocity profile the above basic state can still be regarded as an equilibrium state if viscosity is neglected. Some shear profiles for different widths are illustrated in Fig. \ref{fig:figure01}.\\
%-------------------------------------------------------------------------------------------------------------------------------------------------------------
Our calculations are initialized by adding a small random temperature perturbation to the equilibrium state including the additional shear flow given by equation (\ref{eq:target_shear}). In order to evolve the system in time equations (\ref{eq:NSEquation01}) - (\ref{eq:NSEquations}) are solved by using a hybrid finite-difference/pseudo-spectral code  \cite*[see for example][and references therein]{FLM:340261, 2009MNRAS.400..337S, FLM:8458223, FLM:8885996}.
For the linear stability calculations, the eigenvalue-problem formulated in \cite*{2015AandAWitzke} is numerically solved on a one-dimensional grid in the $z$-direction that is discretised uniformly, and this method is adapted from \citet{2012MNRAS.426.3349F}.\\
%----------------------------------------------------------------------------------------------------------------------------------------------------------
When using direct numerical calculations to solve the system numerically over considerable iterations an issue concerning the momentum conversation might appear, which is specific to the choice of stress-free boundary conditions \cite[see discussion in][]{Jones2011120}. Despite the fact that equation (\ref{eq:NSEquation002}) together with the boundary conditions in equation (\ref{eq:boundary01}) conserves the momentum a cumulative effect of truncation errors at each time step might lead to an unphysical change in momentum when integrating over a vast number of time steps. We have checked that both momentum and mass is indeed conserved for all the calculations presented in this paper.
%----------------------------------------------------------------------------------------------------------------------------------------------------------
%%%%%%%%%%%%%%%%%%%%%%%%%%%%%%%%%%%%%%%%%%%%%%shear_flow_profile%%%%%%%%%%%%%%%%%%%%%%%%%%%%%%
%%%%%%%%%%%%%%%%%%%%%%%%%%%%%%%%%%%%%%%%%%%%%%%%%%%%%%%%%%%%%%%%%%%%%%%%%%%%%%%%%%%%%%%%%%%%%%%%%%
%---------------------------------------------------------------------------------------------------------------------
\subsection{Forcing methods}
\label{sec:Forcing_methos}
We will compare three different methods to sustain an initial shear flow where we distinguish between methods that are static, \textit{i.e.} the force term does not change throughout the calculation, and dynamic forcing methods that might vary depending on the current flow. Furthermore, methods with a force applied to a localised region have a local force whereas for global forcing methods the force term applies on the whole domain. The main goal is to find a forcing method that does not significantly alter the characteristics of the linear phase of the evolution but allows to reach a quasi-steady non-linear state.
\subsubsection{Viscous method}
In order to balance the viscous dissipation associated with the initial shear flow profile given by equation (\ref{eq:target_shear}) the force  
\begin{equation}
\label{eq:viscous_force}
\mathbf{F} = - \sigma C_k \nabla^{2} \mathbf{U}_0
\end{equation}
is added to the RHS of equation (\ref{eq:NSEquation002}). By applying this viscous forcing the initial state given by equations (\ref{eq:equi_temp}) - (\ref{eq:target_shear}) is in equilibrium provided that viscous heating is neglected. This method has been broadly applied in forced shear flows to model the dynamics of the solar tachocline  \citep[e.g.][]{0004-637X-586-1-663, 1538-4357-702-1-L14}. Note that this method only balances for the viscous diffusion of momentum associated with the target profile and does not depend on the actual non-linear solution. In that sense, the forcing can be considered to be local and static.
%--------------------------------------------------------------------------------------------------------------------------------------------------------
\subsubsection{Perturbation method}
Our second method, the perturbation method, was previously used by \cite{1992JFM...237..499H, 1997JFM...342..231J, 2012MNRAS.424..115B}. For this method a slightly different set of differential equations is solved.  A decomposition of the velocity, $\mathbf{u} = \mathbf{U}_0 + \tilde{\mathbf{u}}$ into a background shear flow, $\mathbf{U}_0$, and the deviation from the background profile $\tilde{\mathbf{u}}$ enables the maintenance of a shear flow that is independent of the unstable perturbations. Note, the background velocity profile $U_0$ has to be time independent and divergence free. 
Thus, inserting the above decomposition into the momentum equation (\ref{eq:NSEquation002}) we obtain
\begin{eqnarray}
\rho \frac{\partial}{\partial t} \left( \tilde{\mathbf{u}} +\mathbf{U}_0 \right)& = &  -\,\mathbf{\nabla}p\, + \, \theta(m+1) \rho\, \hat{\mathbf{z}} - \rho \left(\tilde{\mathbf{u}} +  \mathbf{U}_0 \right)  \mathbf{\cdot} \mathbf{\nabla} \left( \tilde{\mathbf{u}} +\mathbf{U}_0  \right)\, \nonumber \\
 &   &  +  \sigma C_k \left[ \nabla^2\left(\tilde{\mathbf{u}} +  \mathbf{U}_0 \right)\, +\,\frac{1}{3}\mathbf{\nabla}\mathbf{\nabla}\mathbf{\cdot}\left(\tilde{\mathbf{u}} +  \mathbf{U}_0 \right)\right],
\end{eqnarray}
since we assume that our background profile does not vary with time and in the flow direction, the time derivative of $\mathbf{U_0}$ and the term $\mathbf{U}_0 \mathbf{\cdot} \mathbf{\nabla} \mathbf{U}_0$ vanish. In order to ensure that the background velocity is not dissipated by viscosity the viscous term associated with it is dropped. The equation takes the form
\begin{eqnarray}
\rho \frac{\partial \tilde{\mathbf{u}}}{\partial t} & = & -\,\mathbf{\nabla}p\, + \, \theta(m+1) \rho\, \hat{\mathbf{z}}  - \rho\tilde{\mathbf{u}}  \mathbf{\cdot} \mathbf{\nabla} \mathbf{U}_0  - \rho   \mathbf{U}_0   \mathbf{\cdot} \mathbf{\nabla} \tilde{\mathbf{u}}\, \nonumber \\
 &   & +\sigma C_k \left( \nabla^2\tilde{\mathbf{u}}\, +\,\frac{1}{3}\mathbf{\nabla}(\mathbf{\nabla}\mathbf{\cdot}\tilde{\mathbf{u}})\right), 
\label{eq:perturbation_method_equation} 
 \end{eqnarray}
such that effectively the differential equation for the velocity perturbations only is solved. Here, the velocity perturbations, $\tilde{\mathbf{u}}$, are initially zero, but the background velocity is incorporated through the two advective terms involving $\mathbf{U_0}$. By the same procedure the equations (\ref{eq:NSEquation01}) and (\ref{eq:NSEquation02}) become:
\begin{eqnarray}
\frac{\partial \rho}{\partial t} & = & - \mathbf{\nabla}\mathbf{\cdot}\left(\rho \tilde{\mathbf{u}} \right) - \mathbf{U_0} \mathbf{\cdot} \mathbf{\nabla} \rho \, \\
\frac{\partial T}{\partial t}  & = & \frac{C_k \sigma (\gamma -1)}{2\rho}|\mathbf{\tau}|^{2}\, + \frac{C_k \sigma (\gamma -1)}{2\rho} \frac{\partial^2 \mathbf{U_0}}{\partial z^2}\, + \, \frac{\gamma C_k}{\rho} \nabla^2 T \nonumber \\
 &  & - \mathbf{\nabla} \mathbf{\cdot}\left(T \, \tilde{\mathbf{u}}\right)\,-\,(\gamma -2)T \mathbf{\nabla} \mathbf{\cdot} \,\tilde{\mathbf{u}} -\mathbf{U_0} \mathbf{\cdot} \mathbf{\nabla} T,
\end{eqnarray}
where the effect of the background velocity, $\mathbf{U_0}$, on the density and temperature is taken into account.
%-------------------------------------------------------------------------------------------------------------------
Mathematically, the equations for $\mathbf{U}_0 + \tilde{\mathbf{u}}$ are the same as for $\mathbf{u}$ in the viscous method. Therefore, both methods should give the same solutions, even if the approach and numerical implementation are significantly different.\\
\subsubsection{Relaxation method}
In the relaxation method, an external force ensures that the flow relaxes towards the initial profile on an arbitrary time-scale $\tau_0$.
We define any quantity $\bar{f}$ as 
\begin{equation}
\label{eq:definition_overbar}
\bar{f}(z)= \frac{1}{N_x N_y} \sum_{i=1}^{N_x} \sum_{j=1}^{N_y} f(i,j,z),
\end{equation}
where the overbar denotes that the quantity $f$ is horizontally averaged, and $N_x$ and $N_y$ are the resolutions in \textit{x}-direction and in \textit{y}-direction respectively. 
The force for the relaxation method in the momentum equation depends on the horizontally averaged velocity $\bar{u}_x (z)$ and is given by
\begin{equation}
\label{eq:relaxation_force}
\mathbf{F}= \frac{\rho}{\tau_0} \left(\mathbf{U}_0 - \bar{u}_x(z) \, \hat{\mathbf{e}}_x \right),
\end{equation}
where $\hat{\mathbf{e}}_x$ is the unit vector in x-direction.
For this method it is crucial to ensure that the forcing is aligned with the flow direction and does not depend on the velocity variation along this horizontal direction, which would otherwise correspond to a local small-scale force. A local forcing should be avoided, because it will suppress any instability and can lead to non-physical behaviour. The relaxation method was used by \citet{refId0m01} to model shear driven turbulence and provides the advantage of an adjustable back reaction on the actual mean flow. It is a global forcing (due to the averaged operator) and a dynamical forcing, because it depends on the actual flow.

%---------------------------------------------------------------------------------------------------------- 
%%
%%-----------------------------------------------------------------------------------------------------------------
%%------------RESULTS
%%------------------------------------------------------------------------------------------------------------------
%%
\section{Comparison of the forcing methods}
\label{Sec:comparison_forceing}
%%%%%%%%%%%%%%%%%%%%%%%%%%%%%%%%%%%%%%%%%%%%%%%%%%%%%%%%%%%%%%%%%%%%%%%%%%%%%%%%%%%%%%%%%%%%%%%%
%%%%%%%%%%%%%%%%%%%%%%%%%%%%%%%%        table_01       %%%%%%%%%%%%%%%%%%%%%%%%%%%%%%%%%%%%%%%%%%%%%%%%
\begin{table}
\centering
\caption{Comparing the linear eigenvalue-solver results with those from non-linear calculations during the linear phase. For case I the shear amplitude is $U_0 = 0.08$ and  $1/L_u = 118$ such that $Ri = 8 \times 10^{-4}$. Taking $C_k = 8 \times 10^{-5}$ results in a  $Pe =34 $.  For case II $U_0 = 0.041$,  $1/L_u = 20$ and $C_k = 1 \times 10 ^{-4}$ such that $Ri = 0.1$ and $Pe =82 $. }

\begin{tabular}{c c c c}
\hline
\textbf{ Method:  }       &   $\zeta_r$   &  $k_{max}$ &  Effective $1/L_u$  \\ 
\hline
\hline
\multicolumn{4}{|c|}{Case I:}   \\
EV-solver  &   $(119 \pm 0.5) \times 10^{-2}$      &  $42.5 \pm 1$   &  118 \\
unforced  &   $( 30 \pm 1)   \times 10^{-2}$      &  $14.1 \pm 1.6$   &  $51 \pm 2 $\\
 1   &  $(110 \pm 5)   \times 10^{-2}$      &  $33.0 \pm 1.6 $   &  $105 \pm 5$ \\
 2   &   $(110 \pm 5)   \times 10^{-2}$      &  $33.0 \pm 1.6$   &  $106 \pm 6$\\
 3 ($\tau_0 = 10 $)   &   $(45  \pm 3) \times 10^{-2}$    &  $14.1 \pm 1.6$   & $37 \pm 2$  \\
 3 ($\tau_0 = 1.0 $)  &  $(80  \pm 3) \times 10^{-2}$    &  $31.4 \pm 1.6$   & $73 \pm 3$  \\
 3 ($\tau_0 = 0.1 $) &   $(105 \pm 5) \times 10^{-2}$    &  $33.0 \pm 1.6$   & $105 \pm 5$   \\
 3 ($\tau_0 = 0.01$)  &   $(110 \pm 5) \times 10^{-2}$    &  $33.0 \pm 1.6$   & $104 \pm 3$   \\ 
\hline
\hline
\multicolumn{4}{|c|}{Case II: }  \\
EV-solver  &  $(53 \pm 0.5) \times 10^{-3}$    & $10.6 \pm 0.1$ &  20\\
unforced   &  $-0.01 $ & not applicable & not applicable \\
 1   &    $(47 \pm 8) \times 10^{-3}$     & $9.4 \pm 1.6$ &  $19 \pm 1$\\
 2   &    $(47 \pm 8) \times 10^{-3}$     & $9.4 \pm 1.6$ &  $19 \pm 1$\\
 3  ($\tau_0 = 0.01$)  &  $(48 \pm 8) \times 10^{-3}$     & $9.4 \pm 1.6$ &  $19.8 \pm 0.3$ \\
\hline
\end{tabular}
\label{table:ParameterSpace}
\end{table}
%%%%%%%%%%%%%%%%%%%%%%%%%%%%%%%%%%%%%%%%%%%%%%%%%%%%%%%%%%%%%%%%%%%%%%%%%%%%%%%%%%%%%%%%%%%%%%%%%%%%%%%%%%
Investigating  saturated flows on long time-scales requires an external force to sustain a target velocity profile.
Here we compare the effects of the different forcings presented in Section \ref{sec:Forcing_methos} on the development of the shear instability. Our investigation is divided into two parts. We compare the linear regime of a shear instability in a two-dimensional framework in Section \ref{sec:results_linear_compare} and the non-linear regime is investigated in Section \ref{sec:3.2non-linear} using three-dimensional calculations. Here we first qualitatively examine two-dimensional slices and three-dimensional renderings of key quantities for different forcing methods in Section \ref{sec:visualisation_comparsion}, before we calculate the horizontally averaged velocity, turbulent Reynolds numbers and turbulent length in Section \ref{subsec:horizontally averaged profiles}. A detailed analysis of the energy budgets is provided in Section \ref{subsec:Energy_budgets}, where the theoretical framework for the energy budgets is introduced in Section \ref{subsec:Energeticconsideration}. The relation between the external work done by the forcing and the amount of dissipated energy by viscosity is analysed in detail in Section \ref{section:work_vs_visousdissipation} and  finally we summarize our findings for the saturated regime in Section \ref{section:discussion_energetics}.
%------------------------------------------------------------------------------------------------------------------------
\subsection{Linear regime}
\label{sec:results_linear_compare}
\begin{figure}
\centering
\includegraphics[width=0.48\textwidth]{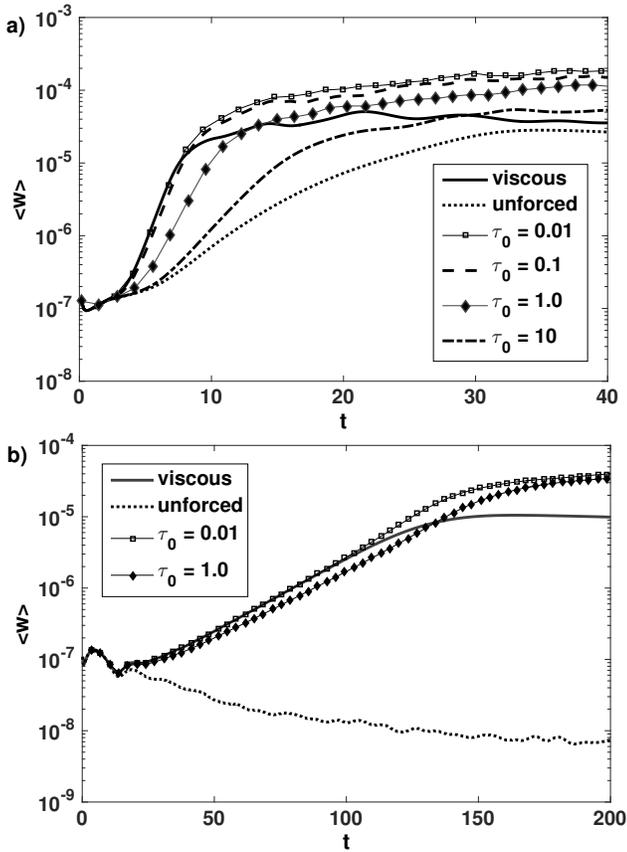}
\caption{The time evolution of the volume averaged vertical velocity, as defined in equation (\ref{eq:definition_angle_bracket}), for the two-dimensional calculations for case I in (a) and  for case II in (b). Different $\tau_0$ parameters for the relaxation method are used and compared to the viscous method and unforced calculations. The vertical velocity is displayed in logarithmic scale and t is given sound-crossing time.}
\label{fig:figure02} 
\end{figure}

Since the initial linear phase of shear flow instabilities is purely two-dimensional, which becomes evident from Squire's theorem \citep{Squire621}, we focus here on calculations in a two-dimensional domain, which has a spatial resolution of $N_x = 512$ and $ N_z = 480$. The stability of a shear flow in a stratified atmosphere is characterized by the non-dimensional Richardson number, $Ri$. Using the general definition of the Brunt-V{\"a}is{\"a}l{\"a} frequency given by
\begin{equation}
N^2(z) =  \frac{g}{\tilde{T}} \frac{\partial \tilde{T}}{\partial z}, \,
\end{equation}
where $\tilde{T} = T (P_{t}/P)^{1-1/\gamma} $ is the potential temperature, the minimum value of the  Richardson number, $Ri$, across the layer is defined as
\begin{eqnarray}
Ri_{min} & = & \min_{0 \le z \le 1} \left(N(z)^2 \left/ \left( \frac{\partial u_0(z)}{\partial z} \right)^2 \right.  \right) \nonumber \\
 & = & \min_{0 \le z \le 1}  \left( \frac{ \theta^2  L_u^2(m +1) \left(\frac{m+1}{\gamma} - m\right)}{\left(1+\theta z \right) \left(U_0-u_0(z)^2/U_0 \right)^2 } \right),
\label{eq:Richardsonnumber_def}
\end{eqnarray} 
where the derivative of the background velocity profile, defined in equation (\ref{eq:target_shear}), with respect to $z$ corresponds to a local turnover rate of the shear. In most cases the minimum $Ri$ value is at $z=0.5$, but for some parameter choices with large temperature gradient, $\theta$, and broad shear width the minimum is shifted towards greater $z$. \\
%----------------------------------------------------------------------------------------------------------
Here we consider unstable shear flows with a Richardson number less than 1/4 at a point in the domain. We do not consider shear instabilities triggered by thermal diffusion for which larger values of $Ri$ can be used \citep{Dudis_1974, Zahn_1974, 1999AA...349.1027L}.  Because the $1/4$ criterion is a necessary, but not sufficient, requirement for instability we also solve the corresponding linear stability problem based on the approach used in \cite*{2015AandAWitzke} in addition to conducting the non-linear calculations.
%-----------------------------------------------------------------------------------------------------------------------
For simplicity, the Prandtl number is fixed to be unity whereas the dimensionless thermal diffusivity $C_k$ is varied from $10^{-4}$ to $10^{-5}$.  
Taking the previous linear study by \citet*{2015AandAWitzke} into account, our parameters satisfy the following requirements: To ensure a stable stratification the polytropic index is set to be $m=1.6$, the amplitude $U_0$ of the shear flow is chosen such that the Mach number in the middle of the domain remains less than $0.08$, which avoids additional stabilisation by compressible effects. Furthermore, we take the initial P\'eclet number, which we define as
\begin{equation}
\label{eq:Peclet_number_in}
Pe = \frac{4 U_0 L_u}{C_k} \, ,
\end{equation}
 to be much greater than unity to avoid a destabilizing effect caused by thermal diffusion. Note, that due to the definition of $U_0$ and $L_u$ in equation (\ref{eq:target_shear}) a factor of $4$ is needed to be consistent with the generally used definition.\\
%--------------------------------------------------------------------------------------------------------------------------------------------------------------
Here we want to compare two linearly unstable cases with distinct behaviour when no external forcing is included. An unforced case will decay if the instability grows on a larger time-scale than the viscous dissipation time-scale. Therefore, we consider two different cases corresponding to two different minimum $Ri$ values. In case I we consider a very small value of $Ri = 8 \times 10^{-4}$, such that the system is unstable even without external forcing. Case II has a greater $Ri = 0.1 $ and greater $C_k$ so that the system is unstable only if we introduce an external forcing. Without forcing, the initial shear flow diffuses quickly such that the Richardson number increases rapidly above 1/4 and no shear instability can be sustained. By considering these two different cases, we investigate how the forcing method used affects the development of a shear instability during the exponential growth phase. Unforced calculations provide a reference for the system's evolution without any external forces. All parameters for case I and case II, the resulting linear growth rates, and the wave number for the most unstable mode for each method are summarised in Table \ref{table:ParameterSpace} using both an eigenvalue solver and direct numerical calculations of unforced and forced cases. \\
The growth rates for the non-linear calculations are obtained by calculating
\begin{equation}
\zeta_r  =  \frac{d \ln{<w>}}{dt} = \frac{1}{<w>} \frac{d<w>}{dt},
\end{equation} 
where the angle brackets denotes that any quantity $f$ is volume averaged as follows
\begin{equation}
\label{eq:definition_angle_bracket}
<f> = \frac{1}{N_x N_y N_z} \sum_{i=1}^{N_x} \sum_{j=1}^{N_y} \sum_{l=1}^{N_z}  f(i,j,l).
\end{equation}
In addition, we fit horizontally-averaged velocity profiles in the shear direction with a hyperbolic tangent profile as in equation (\ref{eq:target_shear}) to estimate the shear width during the exponential regime.  
In order to find the most unstable wave number, which corresponds to the wave number with the most energy, the kinetic energy spectrum is calculated as 
\begin{equation}
E(k_x) = \frac{1}{4}\sum_{k_x} \sum_{z}\hat{\mathbf{u}}(k_x, z) \hat{\rho \mathbf{u}}^*(k_x, z) + \hat{\rho \mathbf{u}}(k_x, z)\hat{\mathbf{u}}^*(k_x, z), 
\end{equation}
where the hat symbol denotes the Fourier transform of the corresponding quantity and the star symbol denotes the complex conjugate.\\
%--------------------------------------------------------------------------------------------------------------------------------------------------------------
The growth rates for the viscous forcing, the perturbation method and the relaxation method (where $\tau_0 = 0.01$) are almost identical for both case I and II. The growth rates achieved by these methods in case II correspond to the growth rate calculated by using the EV-solver with a $12\%$ relative error when taking the growth rate from the EV-solver as reference. For case I the error is $8\%$.
%------------------------------------------------------------------------------------------------------------------------------------------------------------------------
The most unstable mode, $k_{max}$,  in non-linear calculations is the same for the viscous- and the perturbation method. Note that the most unstable wave number obtained by DNS is always slightly smaller than the one obtained from the eigenvalue solver, and more so for case I than for case II.
This is due to a thinner shear width in case I, which is more affected by viscous dissipation, such that the instability is triggered when the shear profile is significantly broader. Therefore, $k_{max}$ deviates for case I more than for case II, where the initial shear width is broader.\\
%%%%%%%%%%%%%%%%%%%%%%%%%%%%%%%%%%%%%%%%%%%%%%%%%%%%%%%%%%%%%%%%%%%%%%%%%%%%%%%%%%%%%%%%%%%%%%%%%%
%Relaxation methods
%%%%%%%%%%%%%%%%%%%%%%%%%%%%%%%%%%%%%%%%%%%%%%%%%%%%%%%%%%%%%%%%%%%%%%%%%%%%%%%%%%%%%%%%%%%%%%%%%%
We now look at the effect of varying the relaxation time-scale $\tau_0$ in the relaxation method. The instability develops always in the middle of the domain and is visually similar to the instability observed by using either the viscous- or the perturbation method. The evolution of $<w>$, for different $\tau_0$ parameters, for case I and case II is shown in Fig. \ref{fig:figure02} and the growth rates of these runs are summarised in Table \ref{table:ParameterSpace}. For all cases the onset of the instability is not sensible to the chosen relaxation time-scale $\tau_0$, but the growth rate decreases with increasing $\tau_0$. This is expected since a smaller $\tau_0$ implies a larger restoring force as soon as the averaged velocity profile differs from the target equation (\ref{eq:target_shear}). Therefore, for relaxation times that are larger the viscous dissipation might be unbalanced such that the initial state changes before an instability is triggered. \\
%-----------------------------------------------------------------------------------------------------------------------------------------------------------------------
The relaxation method leads either to the same instability or sustains an instability triggered by a smoother velocity profile, when the relaxation time $\tau_0$ is increased. To put $\tau_0$ in relation with typical dynamical times the initial turnover time of the shear flow $t_{s} = L_u/U_0$ is calculated, which is $t_s \approx 0.1$ for case I and $t_s \approx 1.2$ for case II. Furthermore, the initial viscous time-scale, $t_{\mu} = L_u^2 /\mu$, that accounts for the time on which the initially shear width is dissipated, gives another reference time. For case I the viscous time-scale is $t_{\mu} = 0.9 $ and for case II it is $t_{\mu} = 25$. Note, that both time-scales represent initial time-scales given by the initial configuration and will change with time as the system evolves, especially in the saturated regime these times might be significant different from the initial values. So that a relaxation time greater than both the viscous and dynamical time-scale corresponds to a very weak back-reaction of the forcing to the change of the averaged velocity. Note that $\tau_0<1$ means that the forcing relaxation is quicker than a sound crossing time which is unlikely to occur in physical system. By varying the relaxation times, $\tau_0$, we investigate how to chose an appropriate $\tau_0$ with respect to the initial time-scales in order to recover the linear instability dictated by the initial state and at the same time a saturated regime that evolves towards a quasi-static state.  \\
%-----------------------------------------------------------------------------------------------------------------------
As expected the numerical calculations using the viscous method and perturbation method lead to exactly the same result, because both methods are mathematically equivalent. Note however that the computational cost is slightly greater when using the perturbation method.
In conclusion the viscous and perturbation method are the same, but the relaxation method shows different dynamics depending on the relaxation time.
If we are to understand which of these forcing methods is most suitable to model shear flows in stellar interiors it is essential to conduct a comparison for the non-linear evolution.
%-----------------------------------------------------------------------------------------------------------------------
%
%-----------------------------------------------------------------------------------------------------
%%%%%%%%%%%%%%%%%%%%%%%%%%%%%%%%%%%%%%%%%%%%%%%%%%%%%%%%%%%%%%%%%%%%%%%%%%%%%%%%%%%%%%%%%%%%%%%%%%%
%---------------------------------------------------------------------------------------------------------------------
%---------------------------------------------------------------------------------------------------------------------
%----------------------------- NON -LINEAR COMPARISON ----------------------------------------------------------------
\subsection{Non-linear phase}
\label{sec:3.2non-linear}
%----------------------------------------------------------------------------------------------------------------------------------------------------------------
%-----------------------------------------------------------------------------------------------------------------------
%---------------  FIGURE:  3dim <w> evolution      ---------------------------------------------------------
%------------------------------------------------------------------------------------------------------------
In order to investigate the non-linear evolution of a stratified shear flow a three-dimensional domain is crucial \citep{JGRC:JGRC3931}, because after a Kelvin-Helmholtz instability 
three-dimensional instabilities are triggered \citep{doi:10.1146/annurev.fluid.35.101101.161144}. These secondary instabilities lead to a turbulent collapse of horizontal vortices, which is suppressed in a two-dimensional setup as discussed by \citet{Scinocca1995}. Therefore, to capture the effect of different forcing methods on the entire dynamics we now focus on three-dimensional calculations, which have a spatial resolution of $N_x = 256$, $ N_y = 256$ and $ N_z = 360$ .\\
%--------------------------------------------------------------------------------------------------------------------------------------
To avoid confinement effects associated with the upper and lower boundaries, we focus on a case with a temperature gradient $\theta = 5$. In this case the instability is more likely to remain confined in a narrow central region as to minimize the importance of our particular choice of boundaries. The polytropic index is kept the same as in the previous section to ensure a stable stratification.  
A parameter search was conducted to find combinations of the other parameters that lead to a finite spread of the unstable region. The dynamical viscosity is taken of order $10^{-4}$ and the Prandtl number, $\sigma = 0.1$, because low Prandtl number flows are more relevant for stellar interiors. The shear flow amplitude is set to $U_0 = 0.2$ and we take $1/L_u = 80$ such that $Ri_{min} = 0.003$. \\
%----------------------------------------------------------------------------------------------------------------------------------------
As discussed above, the viscous method and the perturbation method are equivalent, such that in the following, only results obtained with the viscous method are shown but we have checked that the calculations are indeed the same when using the perturbation method. Our study compares the resulting non-linear dynamics obtained by using the viscous method and the relaxation method. Furthermore, we discuss the effect of varying $\tau_0$ in the relaxation method and how does it compare with the viscous method, taken as reference. \\
%-------------------------------------------------------------------------------------------------------------------

\begin{figure*}
  \vspace*{5pt}
\includegraphics[width=0.45\textwidth]{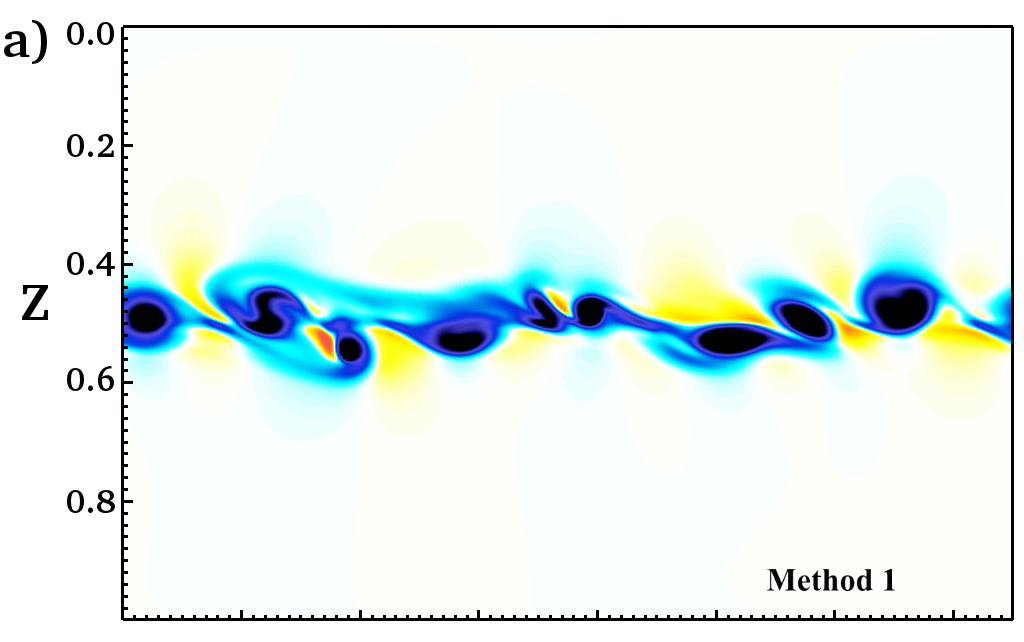}
\includegraphics[width=0.42\textwidth]{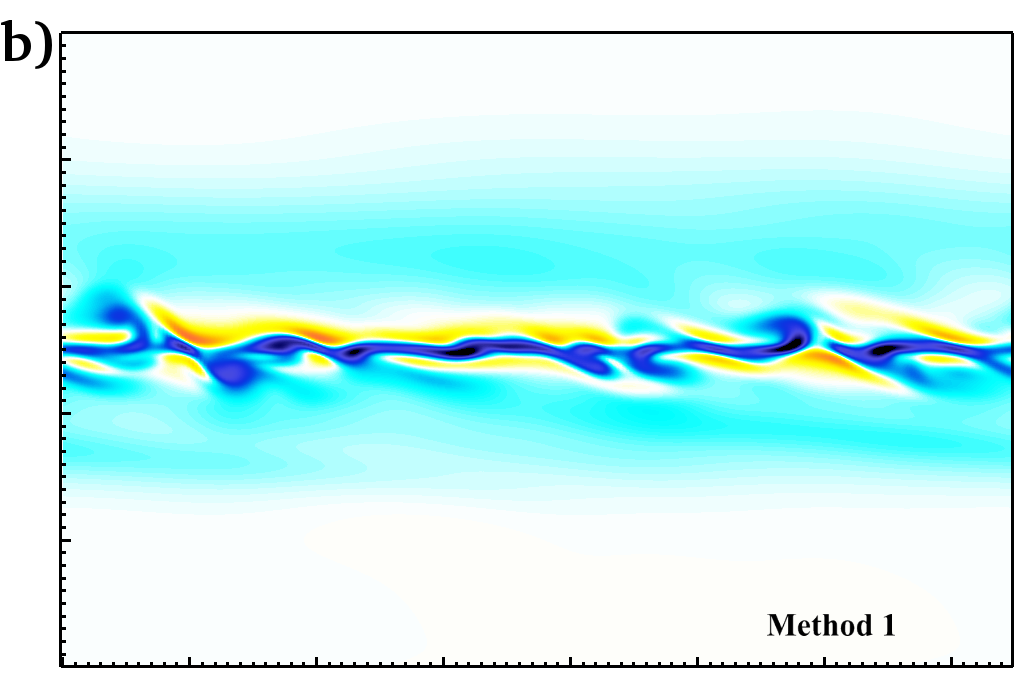}
\includegraphics[width=0.08\textwidth]{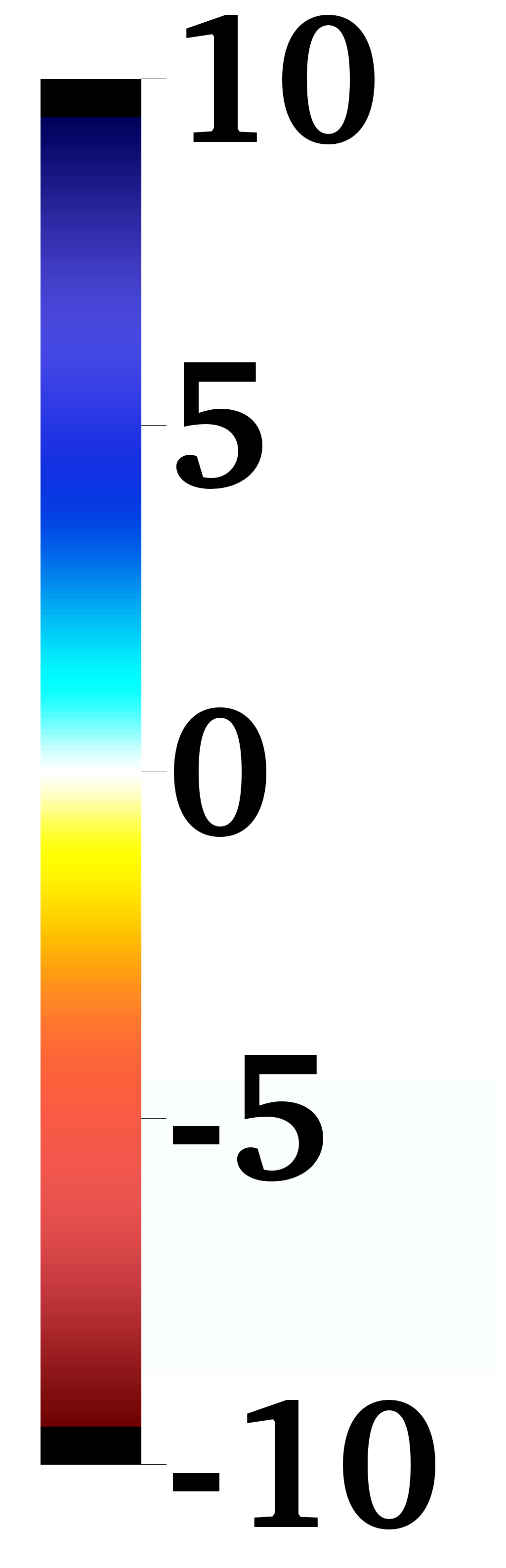}
\includegraphics[width=0.45\textwidth]{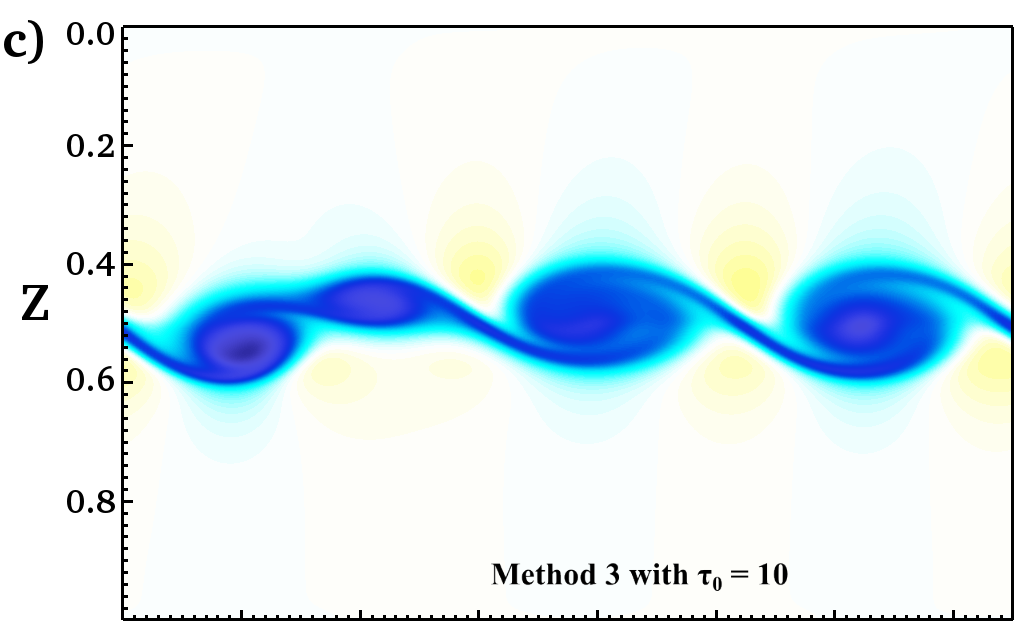}
\includegraphics[width=0.42\textwidth]{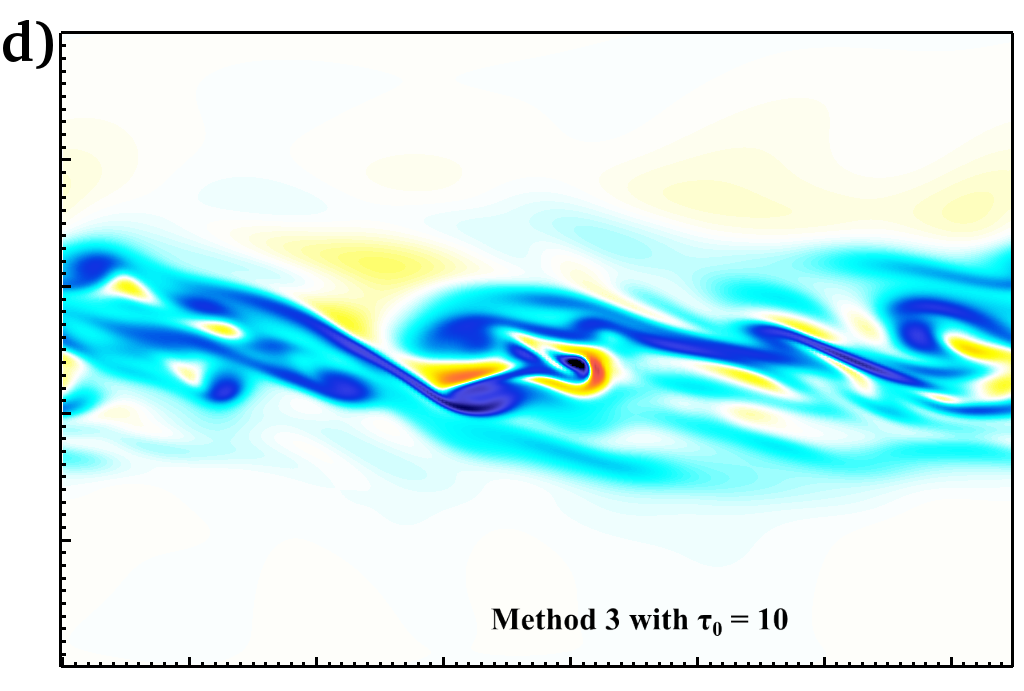}
\includegraphics[width=0.08\textwidth]{./figures/coulorscale0000.png}
\includegraphics[width=0.45\textwidth]{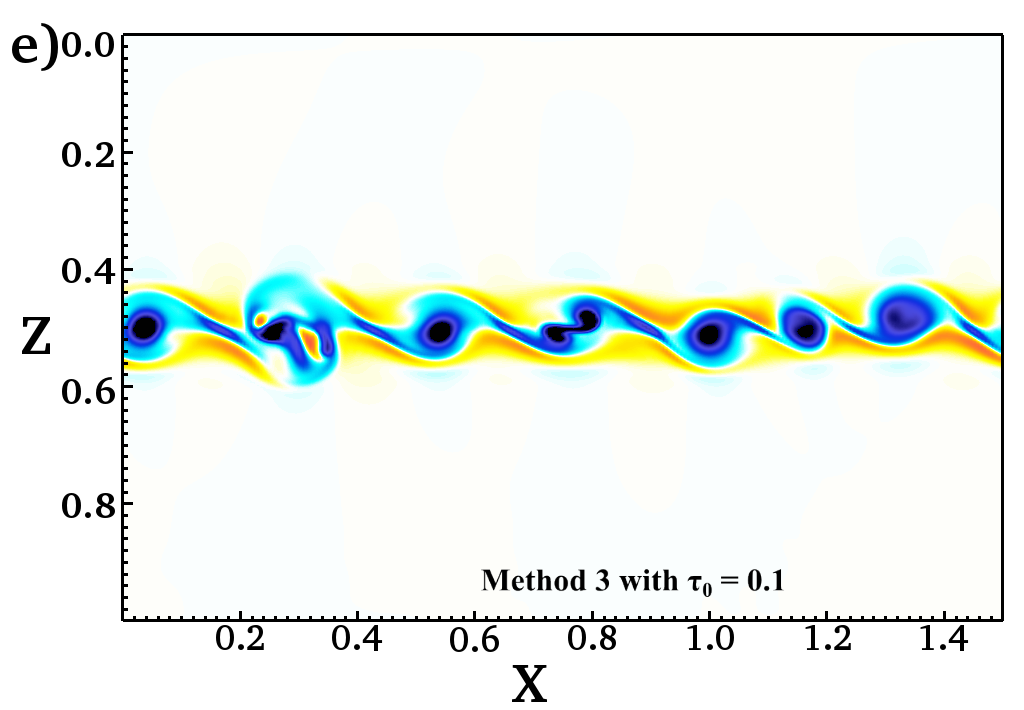}
\includegraphics[width=0.42\textwidth]{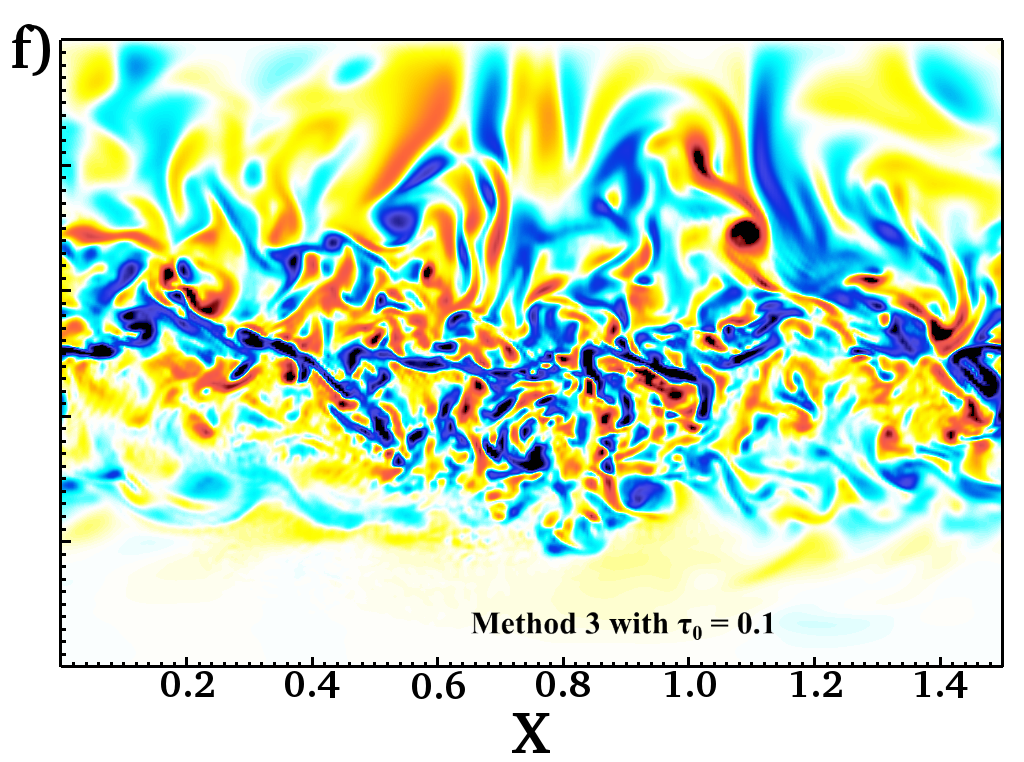}
\includegraphics[width=0.08\textwidth]{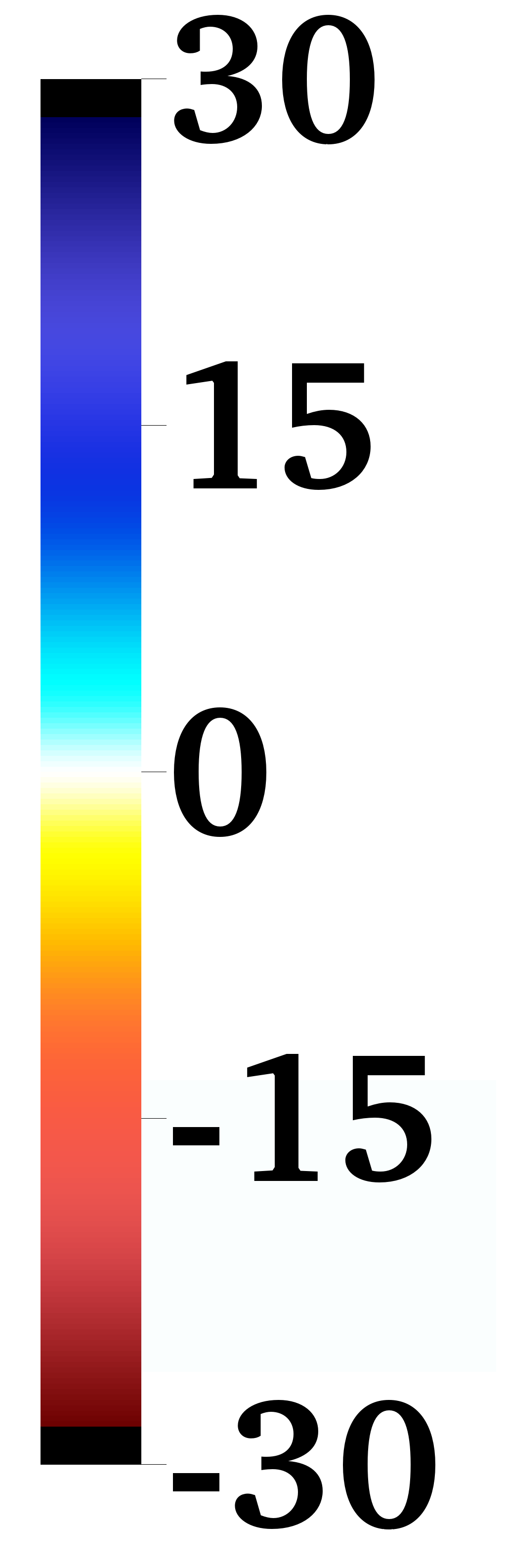}
\caption{The vorticity component perpendicular to the x-z-plane for different forcing methods at two different stages during the time evolution. The plots at the top (a), (b) show snapshots of two different times for the viscous method, where (a) is at $\tilde{t} = 7.2$, (b) at $\tilde{t} = 40$. The middle row (c) and (d) show the relaxation method with $\tau_0 =10$ at $\tilde{t} = 6.8$ and $\tilde{t} = 40$, respectively.  In (e) and (f) the relaxation method with $\tau_0 = 0.1$ was used,  where $\tilde{t} = 12.3$ in (e) and $\tilde{t} = 38$ in (f).}   
\label{fig:figure03} 
\end{figure*}
%-------------------------------------------------------------------------------------------------------------------
%
%         Visualisation (3.2.1)
%
%-------------------------------------------------------------------------------------------------------------------
\subsubsection{Visualisation}
\label{sec:visualisation_comparsion}

%-------------------------------------------------------------------------------------------------------------------------
\begin{figure}
%Figure05
\centering
\includegraphics[width=0.40\textwidth]{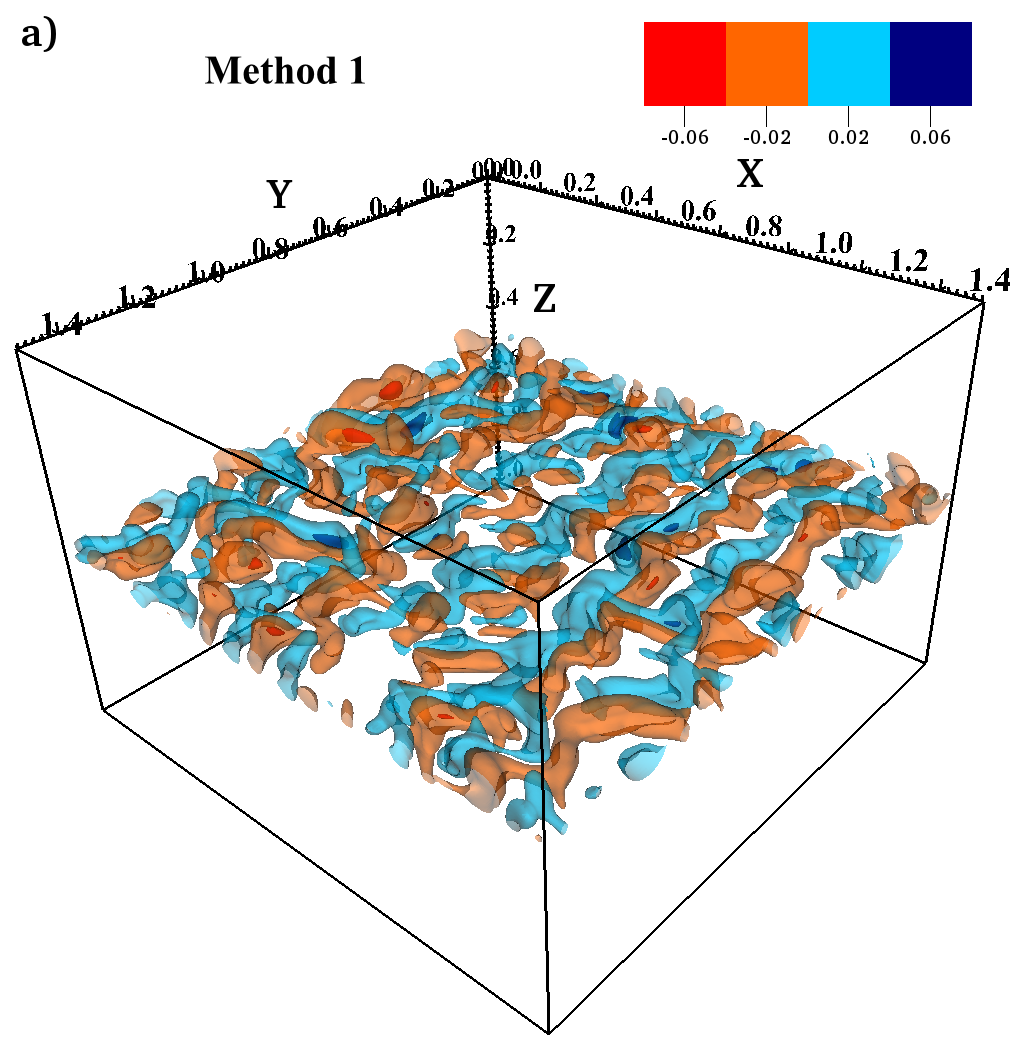}
\includegraphics[width=0.40\textwidth]{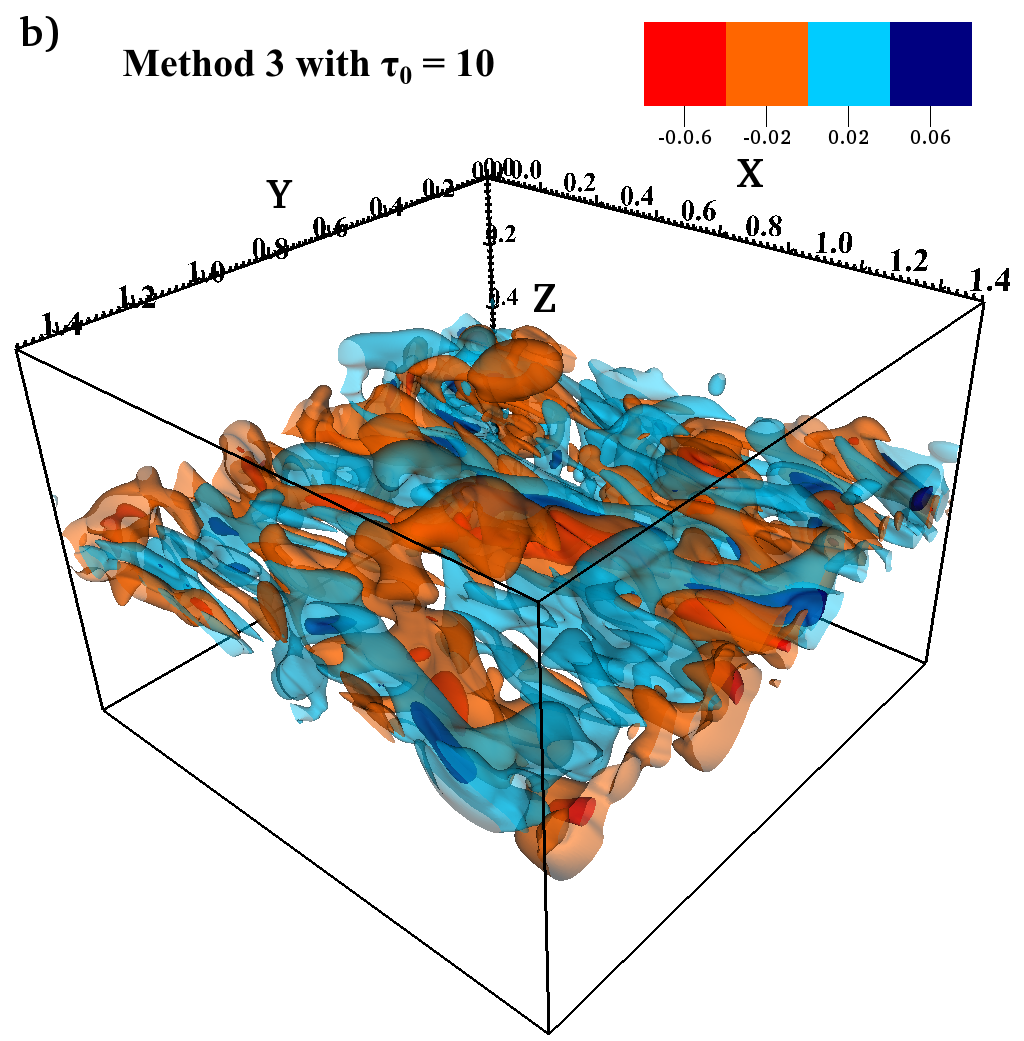}
\includegraphics[width=0.40\textwidth]{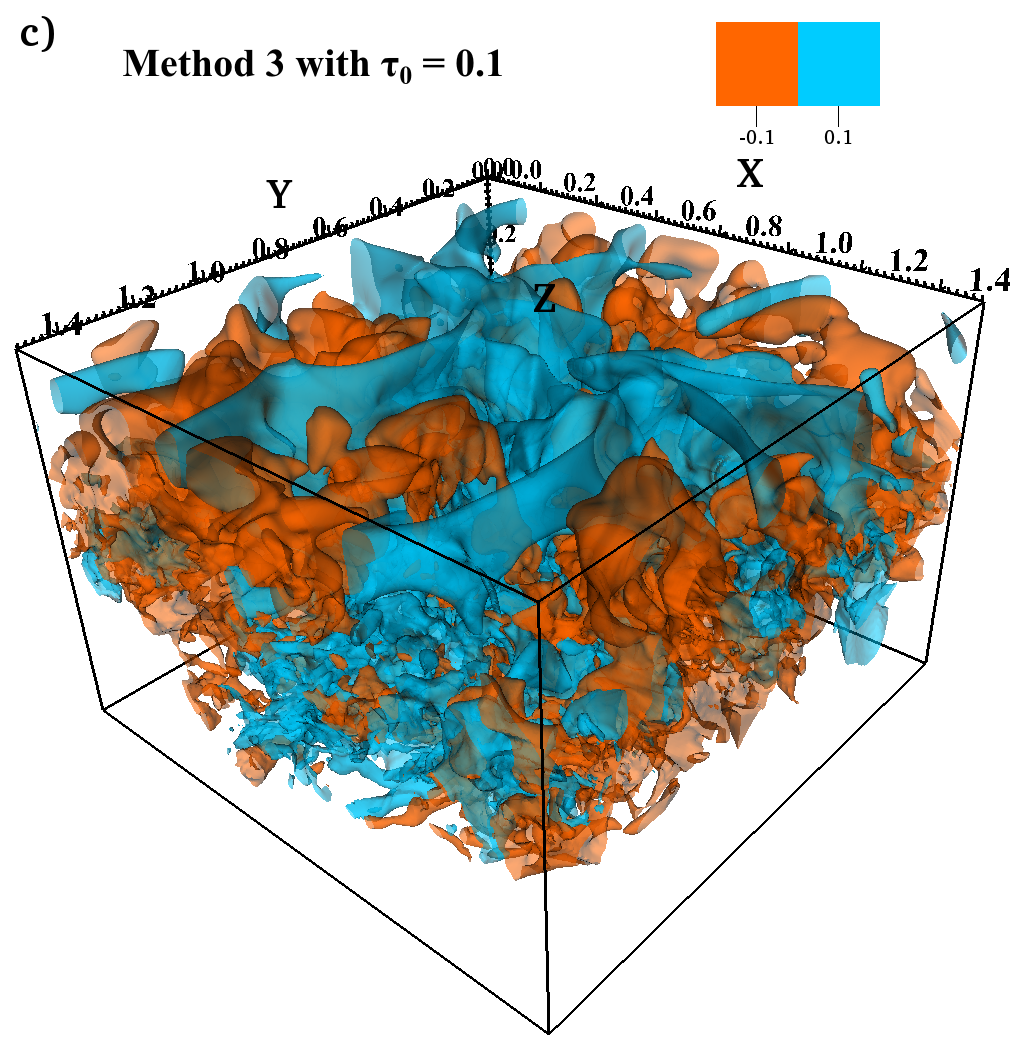}
\caption{The vertical velocity component w for three different forcing methods at several sound crossing times after saturation. In (a) the viscous method is used at $\tilde{t} \approx 40$. In (b) the relaxation method is used with $\tau_0 =10 $ at $\tilde{t} \approx 40$ and in (c) the averaged method is used with $\tau_0 = 0.1$ at $\tilde{t} \approx 38$.}   
\label{fig:figure04} 
\end{figure}
In order to compare the flow evolution we start with a visualization of the vorticity at three different stages during the non-linear saturation.
The first stage is the exponential growth phase, during which all forcing methods show a similar evolution, where the layer with non-zero vorticity spreads vertically.  For the viscous forcing and the relaxation method with $\tau_0 \ll 1$, this layer remains significantly smaller than for calculations where the relaxation method with $\tau_0 \sim \mathcal{O}(1)$ is used. Note that the the viscous time-scale for this case is $t_{\mu} \approx 1.5$.\\
%--------------------------------------------------------------------------------------------------------------------------------------------------------------------------------------------------
The second stage is chosen at a point when the instability starts to saturate and the fluid parcels overturn. In Fig. \ref{fig:figure03} the vorticity component perpendicular to the x-z plane for the viscous forcing, the relaxation method with $\tau_0 = 0.1$ and $\tau_0 = 10$ for the second and third stage are plotted in the x-z plane at y=0.8. For the second stage the dynamics differ between the different forcing methods, which can be seen in Fig. \ref{fig:figure03} (a), (c) and (e). In  Fig. \ref{fig:figure03} (a) small patches of strong positive vorticity are merging together into each other along a thin horizontal layer and a few small negative vorticity patches are present. In comparison, when using the relaxation method with $\tau_0 = 10$ large billows of smaller positive vorticity occupy a horizontal layer which is more extended in the vertical direction, see Fig. \ref{fig:figure03} (c). This can be explained by the smoother shear width, which is a consequence of the slow back-reaction of the forcing. Using a smaller $\tau_0$ leads to a greater vorticity amplitude than achieved by the viscous method (see Fig. \ref{fig:figure03} (a) and (e), note the different color scales) while the spread of the instability remains similar.\\
%--------------------------------------------------------------------------------------------------------------------------------------------------------------------------------------------------
The third stage for the different methods is several sound crossing times after saturation, where the flow is evolving towards a quasi-static state. Comparing Fig. \ref{fig:figure03} (b), where the viscous method is used with Fig. \ref{fig:figure03} (d), and Fig. \ref{fig:figure03} (e), where the relaxation method with $\tau_0 = 10$ and $\tau_0 = 0.1$ is used respectively, the main differences are the vertical extent of the overturning region and the amplitude of the vorticity.
Using a larger $\tau_0 = 10$ leads to a similar situation as for the viscous method which becomes evident when comparing Fig. \ref{fig:figure03} (b) and (d). In Fig. \ref{fig:figure03} (b) the layer is thin and shows elongated regions of strong positive vorticity, whereas in  Fig. \ref{fig:figure03} (d) the region is significantly extended with small patches of strong positive and negative vorticity. A few larger regions are present further away from the middle of the domain. For the relaxation method with $\tau_0 = 0.1$ a drastically different behaviour is observed, see Fig. \ref{fig:figure03} (f), where the vorticity amplitude is much greater and the region where overturning is present is taking up almost the entire domain. In addition, much more small scale turbulence occurs around $z=0.5$.  This can be explained by the form of the forcing used: The viscous method acts with a force, that is confined in a narrow region around the middle of the domain such that the instability can develop freely further away from $z=0.5$. The relaxation method drives the fluid towards the target profile even far away from the middle of the domain. This supports additional spread of the shear instability and triggers more turbulent motion. While it is worth mentioning that for the relaxation method with $\tau_0 = 0.1$ due to strong viscous heating convective motion is present in the upper half of the domain (which we checked by calculating the Brunt-V{\"a}is{\"a}l{\"a} frequency), this dynamics will not be further discussed, as it is the result of the artificially low value of $\tau_0$ used in that case. Furthermore, we expect this unrealistic instability to disappear at lower Prandtl numbers.\\
%--------------------------------------------------------------------------------------------------------------------------------------------------------------------------------------------------
Fig. \ref{fig:figure04} shows contour plots of the vertical velocity in three dimensions at approximately 40 sound-crossing times, which is well after the non-linear saturation. For the viscous method the patches of downwards and upwards motion form an alternating pattern along the x-direction where regions of the same velocity are arranged in small tubes that are extended along the y-direction, see Fig. \ref{fig:figure04} (a). Such a pattern is not present in Fig. \ref{fig:figure04} (b), where the relaxation method with $\tau_0 = 10$ is used. Here, the regions of the same velocity form larger patches which are extended along the x-direction and changes sign along the y-direction.  This indicates that a secondary instability which forms overturning billows along the y-direction is more dominating when the relaxation method with larger values of $\tau_0$ is used. For the relaxation method with $\tau_0 = 0.1$, displayed in Fig. \ref{fig:figure04} (c), the artificially strong forcing leads to an intense forward energy cascade and associated small-scale turbulent motions in the middle of the layer. The large-scale structures observed in the upper part of the domain are convective cells resulting from the large central viscous heating.\\
%--------------------------------------------------------------------------------------------------------------------------------------------------------------------------------------------------
When looking at the time evolution along Fig. \ref{fig:figure03} (a), (b)  and the evolution along Fig. \ref{fig:figure03} (e), (f) a significant difference in the amplitude of the vorticity can be noticed. While for the viscous method the amplitude increases towards a peak during the saturation, Fig. \ref{fig:figure03} (a), and start to decrease afterwards, for the relaxation method the amplitude of the vorticity constantly increases and reaches a maximum after saturation, see Fig. \ref{fig:figure03} (b) and (f). This might be explained by the form of the forcing: Because the relaxation method adjusts the magnitude of the force according to the mean flow, the  strength of the force increases constantly and lead to more overturning with time, whereas the force remains constant when using the viscous method, such that the overturning settles down after saturation. However, in order to check if this is indeed the case a more detailed analysis on the work done by the force and the total viscous dissipation is required, which is discussed in Section \ref{subsec:Energy_budgets}.\\
Having compared the non-linear evolution for different forcing methods qualitatively we can conclude the following. The viscous method and the relaxation method with $\tau_0 \sim  \mathcal{O}(t_{\mu})$ or larger result in similar, but still distinct, evolutions. Using a relaxation time $\tau_0 \ll t_{\mu}$ but still greater than the dynamic time-scale $t_s \approx 0.06$ leads to a very different non-linear dynamics with great mixing and possible non-physical behaviour leading to convection. Therefore, we conclude that, for the saturated regime, the initial viscous time-scale, here $t_{\mu} \approx 1.5$, gives a reference time for the choice of
 $\tau_0$ and the case with $\tau_0 = 0.1$ will be excluded from further analysis.
%
%-----------------------------------------------------------------------------------------------------------------------------------------
%
%    Horizontally averaged profiles (3.2.3)
%
%-----------------------------------------------------------------------------------------------------------------------------------------
\subsubsection{Horizontally averaged profiles}
\label{subsec:horizontally averaged profiles}
%-----------------------------------------------------------------------------------------------------------------------------------------
% turbulent Reynolds number
%-----------------------------------------------------------------------------------------------------------------------------------------
\begin{figure}
  \vspace*{5pt}
\includegraphics[width=0.475\textwidth]{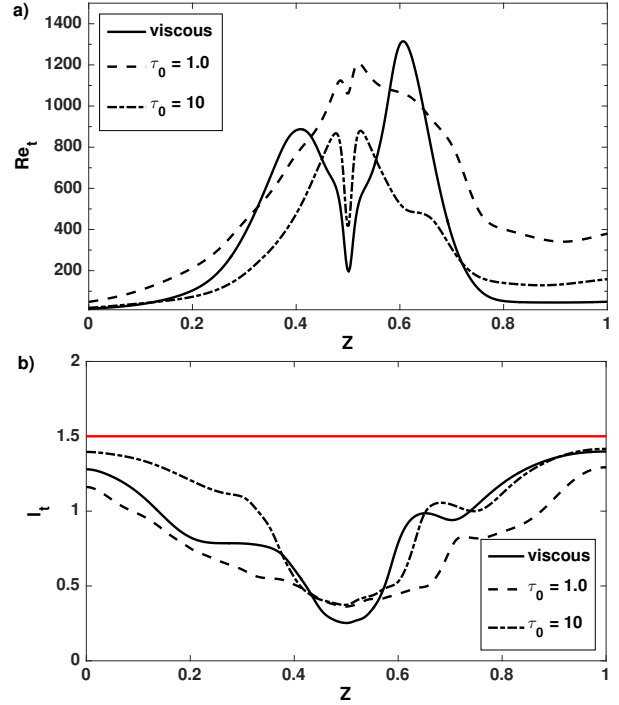}
\caption{The turbulent Reynolds number and turbulent length-scales for three different forcings. In (a) $Re_t$, as defined in equation (\ref{eq:turbulent_reynolds_number}), is plotted versus z and (b) shows $l_t$, where the red line indicates the horizontal extend of the domain. }   
\label{fig:figure08} 
\end{figure}
\begin{figure}
\centering
\includegraphics[width=0.475\textwidth]{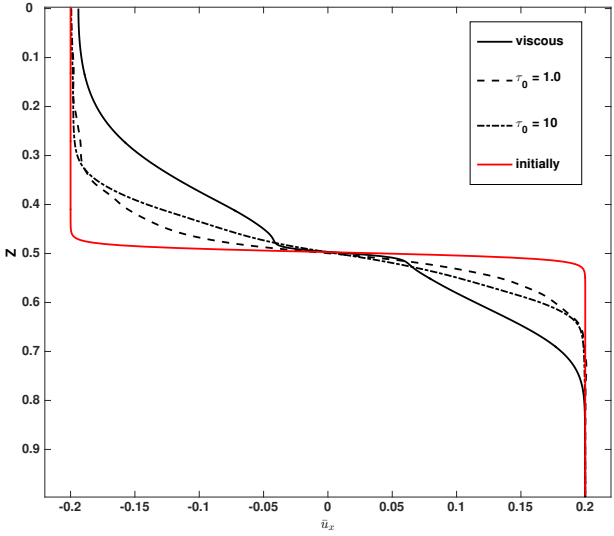}
\caption{The horizontally averaged $u_x$ profiles at $t \approx 60$ are shown for different forcing methods. Two different $\tau_0$ parameters are considered for comparison.} 
\label{fig:figure09}
\end{figure}

%-----------------------------------------------------------------------------------------------------------------------------------------%-----------------------------------------------------------------------------------------------------------------------------------------
The system under consideration is stratified, such that most quantities will change with depth, \textit{z}, throughout the domain. Therefore, investigating horizontally averaged profiles with depth provides further insight in the system dynamics during the saturated regime. \\
%---------------------------------turbulent Reynolds number----------------------------------------------
Let us first define a local turbulent Reynolds number as follows
\begin{equation}
\label{eq:turbulent_reynolds_number}
Re_{t} (z) = \bar{\rho}(z) l_t(z) u_{rms}(z) / (\sigma C_k),
\end{equation}
where $\bar{\rho}(z)$ is the horizontally averaged density.  Here $u_{rms}$(z) is the root mean square of the fluctuating velocity averaged over the horizontal layers calculated as follows
\begin{equation}
u_{rms}(z) = \frac{1}{ N_x N_y} \sum_{x =1}^{N_x} \sum_{y = 1}^{N_y} \sqrt{\left(\mathbf{u}(x,y,z) - \mathbf{U_0}(z)\right)^2},
\end{equation}
where $\mathbf{U_0}(z)$ is the target velocity profile as defined in equation (\ref{eq:target_shear}).
The turbulent length-scale is taken as
\begin{equation}
\label{eq:turbulent_length}
l_t(z) = 2 \pi \frac{\int E(k,z)/k \, dk}{ \int E(k,z)\, dk }, 
\end{equation} 
where $k^2 = k_x^2 + k_y^2$  is the horizontal wave number and the energy spectrum $E(k,z)$ is averaged over horizontal layers such that it takes the form 
\begin{equation}
E(k,z) = \frac{1}{4}\sum_{k} \hat{\mathbf{u}}(k_x, k_y, z) \hat{\rho \mathbf{u}}^*(k_x, k_y, z) + \hat{\rho \mathbf{u}}(k_x, k_y, z)\hat{\mathbf{u}}^*(k_x, k_y, z).
\end{equation} 
The resulting Reynolds number variations with depth for the four different runs after 60 sound-crossing times is shown in Fig. \ref{fig:figure08} (a). It becomes evident that the flow for all four cases is very different at the late stage of the calculation. The viscous method is characterised by two regions above and below the middle of the domain, where the flow has high Reynolds numbers about 850 and 1300 respectively. The asymmetry is due to significantly denser fluid at the bottom of the domain. For the relaxation method with $\tau_0 =10$, these two regions are narrower and the maximal Reynold numbers are approximately 800 and 900. For $\tau_0 = 1.0$ the region of high $Re_t$ is spread with two peaks very close to $z=0.5$. For these three methods a moderate turbulent flow confined in the middle is obtained. The corresponding turbulent length-scales, shown in Fig. \ref{fig:figure08} (b), reveal that  for the three methods used the length-scales become less than $0.5$ around $z=0.5$  and increase towrads the boundaries. Using the viscous method leads for this particular case to smaller turbulent length scales than using a relaxation method. \\
%---------------------------------------------------------------------------------------------------------------------------------------------------------------
We now consider the horizontally averaged velocity profile $\bar{u_x}(z)$ shown in Fig. \ref{fig:figure09} at a time, $\tilde{t}$, where the system evolves towards a quasi-static state. For all methods the shear profile is smoothed out. For the relaxation method $\bar{u_x}(z)$ remains a hyperbolic tangent profile, but for the viscous method a steep transition occurs around $z= 0.5$, which merges into a smoother region at the boundaries. This is caused by the different type of forcing: The force applied in the viscous method is solely defined by the shape of the target profile. Since the initial target profile has a thin width, the second derivative is large in this region, which causes a stronger forcing, compared to the outer parts of the domain. In contrast the relaxation method applies a force that depends on the deviation of the actual mean profile from the target profile, such a back reaction ensures the preservation of a hyperbolic tangent profile.\\
%----------------------------------------------------------------------------------------------------------------------------------------------------------------
%-----------------------------------------------------------------------------------------------------------------------
%---------------  FIGURE:  horizontally averaged u_x      ---------------------------------------------------------
%------------------------------------------------------------------------------------------------------------
%-----------------------------------------------------------------------------------------------------------------------------------------
%
%    Energetic considerations  (3.2.2)
%
%-----------------------------------------------------------------------------------------------------------------------------------------
%
%-------------------------------------------------------------------------------------------------------------------
\subsubsection{Theoretical framework for energy budgets}
\label{subsec:Energeticconsideration}
To get a more comprehensive insight on the processes involved when a linear shear flow instability saturates it is useful to track the evolution of the standard forms of energy during the transition phase and beyond. Following the same procedure as used in \citet{landau2013fluid} and \citet{Griffies_climate} we decompose the total energy into the kinetic energy, $E_{kin}$,  internal energy, $I$, and gravitational potential energy, $E_{pot}$, which in our case are given as
\begin{eqnarray}
\label{eq:kin_energy}
 E_{kin} & = & \frac{1}{2} \int_V \rho \mathbf{u}^2 dV \\
 \label{eq:internal_energy}
I & = & c_v \int _V T \rho dV\\
\label{eq:potential_energy} 
 E_{pot} & = &-\theta(m+1) \int_V \rho z dV,
 \end{eqnarray}
where the volume integral is taken over the whole domain and the internal energy for an ideal gas is considered.
Then, in order to understand the energy evolution with time, and to get more detailed insights into energy budgets, it is useful to derive the energy changes in our system using equations (\ref{eq:NSEquation01}) - (\ref{eq:NSEquation02}). The time derivative of the kinetic energy becomes
\begin{eqnarray}
\label{eqn:kin_energy_changes}
\frac{\partial E_{kin}}{\partial t} & = & \frac{\partial }{\partial t} \left(\frac{1}{2} \int_V{ \mathbf{u \cdot u} \, \rho  \, dV} \right) \\ \nonumber
%
%& = & \int_V{\mathbf{u \cdot} \frac{\partial}{\partial t}\left(\rho \mathbf{u}\right)  \, dV} - \frac{1}{2} \int_V{\mathbf{u \cdot u} \, \frac{\partial}{ \partial t}\rho \, dV} \\ \nonumber
%
& = & \underbrace{\sigma C_k \oint_S{\mathbf{\tau \cdot u} \mathbf{\cdot \hat{n}}\, dS}}_{=0} - \underbrace{\sigma C_k \int_V{\tau_{i j} \frac{\partial u_i}{\partial x_j} dV}}_{\varepsilon} \\ \nonumber
& & - \underbrace{\frac{1}{2}\oint_S{|\mathbf{u}|^2 \rho \mathbf{u \cdot \hat{n}}  dS}}_{\mathcal{H}_{a}} - \underbrace{\int_V{\mathbf{u} \mathbf{\cdot} \mathbf{\nabla} p dV}}_{\mathcal{H}_{p}}\\ \nonumber
%&  &  + \underbrace{   \frac{1}{2} \int_{V}{ \mathbf{u \cdot u} \, \mathbf{\nabla \cdot (\rho \mathbf{u})} dV} - \int_V{\mathbf{u \cdot} \mathbf{\nabla} \mathbf{\cdot} \left(\rho \mathbf{u u} \right) dV}}_{\mathcal{H}_{a}} \\ \nonumber
%
&  &   +  \underbrace{ \int_V{\theta(m+1) \rho w \, dV}}_{\mathcal{H}_{\rho}} + \underbrace{\int_V \mathbf{u} \cdot \mathbf{F} dV}_{W} \\ \nonumber
& = & - \varepsilon - \mathcal{H}_{a} - \mathcal{H}_{p} +  \mathcal{H}_{\rho}  + W
\end{eqnarray}
where $S$ denotes the volume surface,  $\varepsilon$ is the positive viscous dissipation rate, $\mathcal{H}_{a} $ is the change rate due to advection, $\mathcal{H}_{p}$ is the rate of work done by expansion and contraction, $\mathcal{H}_{\rho}$ denotes the exchange rate with the potential energy due to density flux and $W$ is the work done by external forcing.
For the rate of change in the internal energy we get
\begin{eqnarray}
\label{eqn:internal_energy_changes}
\frac{\partial}{\partial t} I  & = &  \frac{\partial}{\partial t} \left( c_v \int_V{T\rho dV}\right)\\ \nonumber
% & = &  c_v \int_V{\partial_t(T) \rho dV}  + c_v \int_V{T \partial_t(\rho) dV} \\ \nonumber
%
& = &   \underbrace{ c_v \oint_S{\gamma C_k \nabla T dS} }_{\Phi_{temp} } + c_v (\gamma -1) \underbrace{\int_V{ \mathbf{u} \mathbf{\cdot} \mathbf{\nabla} p dV}}_{\mathcal{H}_{p} }  +\underbrace{ c_v  \int_V{ \rho q dV}}_{\varepsilon } \\ \nonumber
& = &  \Phi_{temp} + \mathcal{H}_{p} + \varepsilon, 
\end{eqnarray} 
where $q =  C_k \sigma (\gamma -1)|\mathbf{\tau}|^{2} /{2\rho} $ such that $\varepsilon$ is due to viscous heating and $\Phi_{temp}$ corresponds to heat loss or gain at the surface, $S$.    
In the standard form, the changes in the gravitational potential energy are only due to the exchange of density flux as can be seen by taking the time derivative 
\begin{eqnarray}
\label{eqn:grav_pot_energy_changes}
\frac{\partial}{\partial t} E_{pot} & = & \frac{\partial}{\partial t}  \left( - \int_V \theta(m+1) \rho z dV \right)\\ \nonumber
% & = & \theta(m+1) \int_V{z \mathbf{\nabla \cdot}(\rho \mathbf{u}) }\\ \nonumber
& = &  \underbrace{\theta(m+1) \oint_S{ z \rho \mathbf{u} \mathbf{\cdot} \mathbf{\hat{n}} dS}}_{=0}
  - \underbrace{\theta(m+1) \int_V{ \rho w  dV}}_{\mathcal{H}_{\rho} } = -\mathcal{H}_{\rho}.
\end{eqnarray}
Summing equations (\ref{eqn:kin_energy_changes}) - (\ref{eqn:grav_pot_energy_changes}) yields the total energy change of the system
\begin{equation}
\frac{\partial}{\partial t} \left(E_{kin} + E_{pot} + I \right) = W + \Phi_{temp} -\mathcal{H}_{a},
\end{equation}
which is only due to external forces, heat loss or gain at the surfaces, and advection. Note, that $\mathcal{H}_{a}$ is negligible in our case, because our system is closed and mass is conserved. It can be seen that viscous dissipation, $\varepsilon$, density flux,$\mathcal{H}_{\rho}$, and work done by expansion and contraction, $\mathcal{H}_{p}$ are exchanged between the kinetic energy and the potential energies. However, using the standard decomposition of energies it remains impossible to distinguish between reversible and irreversible energy exchange between these three energy budgets. In order to resolve this issue
\cite{FLM:354020} introduced a method to analyse the mixing behaviour of a turbulent flow, which can be used to track reversible and irreversible changes of different potential energies. This framework was further extended to compressible fluids by \cite{1402-4896-2013-T155-014033}.
For this method we decompose the gravitational potential energy of the system defined in equation (\ref{eq:potential_energy}) into two parts.
One part is the so-called background potential energy defined as
\begin{equation}
E_{back}= -\theta(m+1) \int_V \rho_{\star} z dV,
\end{equation}
where the $\rho_{\star}$ is the adiabatically redistributed density. This definition is also appropriate for a compressible fluid. Another part is the available potential energy  
\begin{equation}
E_{avail}= -\theta(m+1) \int_V \left(\rho- \rho_{\star}\right) z dV,
\end{equation}
which corresponds to the difference between the actual potential energy $E_{pot}$ and $E_{back}$.
The available potential energy can be transformed into other types of energies, whereas the background energy can not be accessed and transformed in other types of energies. Therefore, the background potential energy can be interpreted as the part of the total gravitational potential energy that corresponds to the lowest energy level that can be reached if the system is adiabatically redistributed. While a system is evolving the background potential energy can be only changed by irreversible processes. In our numerical calculations the background potential energy is obtained by taking the actual density distribution and sorting it in an ascending order, such that the highest density is at the bottom of the domain. In a similar procedure the internal energy can be decomposed into a background internal energy budget and an available internal energy budget, for a more detailed discussion see \citet{1402-4896-2013-T155-014033}. However, for our purpose it is sufficient to analyse only the budgets for the gravitational potential energy in order to understand the mixing behaviour of the system.\\
%----------------------------------------------------------------------------------------------------------------------------------------
%
\subsubsection{Energy budgets from numerical calculations}
\label{subsec:Energy_budgets}
%------------------------------------------------------------------------------------------------------------------------------------------
\begin{figure}
  \vspace*{5pt}
\includegraphics[width=0.475\textwidth]{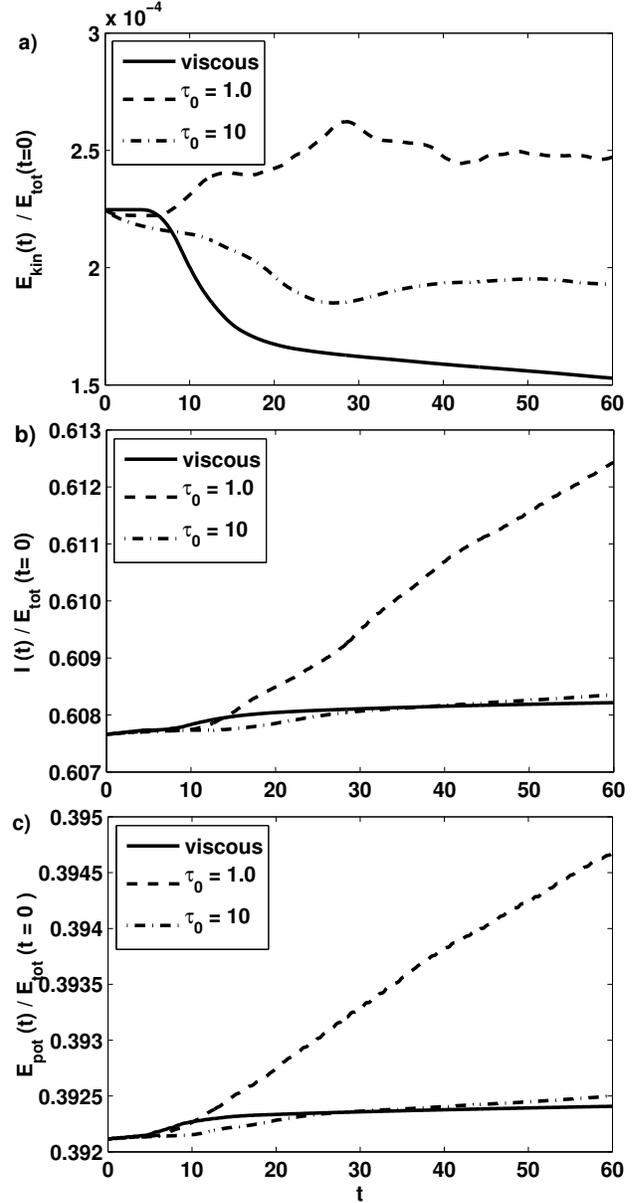}  
\caption{The total kinetic energy, internal energy and the gravitational potential energy evolution for viscous forcing and relaxation method by using two different $\tau_0$. In (a) $E_{kin}(t)/ E_{tot}(t = 0)$ is plotted with time, in (b) $I(t) / E_{tot}(t=0)$ and in (c) $E_{pot} / E_{tot} (t=0)$ is displayed, where all energies are normalised by the initial total amount of the system's energy  $E_{tot}(t=0) = 195.81$.
}   
\label{fig:figure05} 
\end{figure}
%-------------------------------------------------------------------------------------------------------------------
%---------------------------------------------------------------------------------------------------------------------------------------- 
%
%----------------------------------------------------------------------------------------------------------------
%-----------------------------------------------------------------------------------------------------------
\begin{figure}
  \vspace*{5pt}
\includegraphics[width=0.475\textwidth]{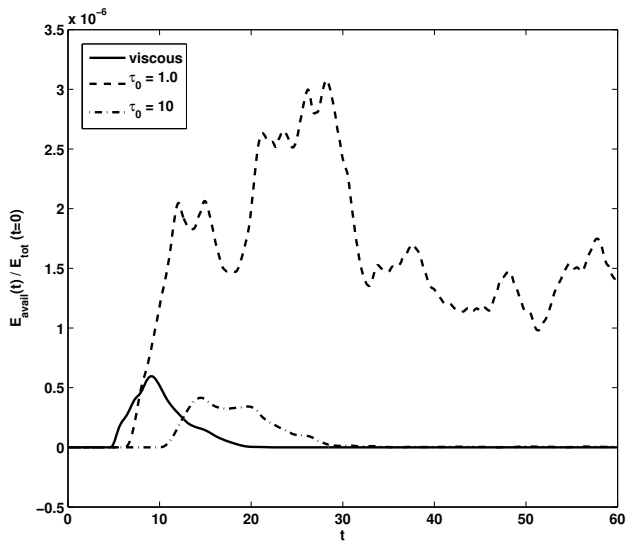}
\caption{The available gravitational potential energy $E_{avail} (t) / E_{tot}(t = 0)$  evolution for both the viscous forcing and relaxation methods by using two different $\tau_0$.
}   
\label{fig:figure06} 
\end{figure}
%------------------------------------------------------------------------------------------------------------------- 
\begin{table}
\centering
\caption{Time intervals of the different dynamical stages for the viscous forcing and relaxation method with two different $\tau_0$.} 
\begin{tabular}{c c c c}
\hline
\hline
                         &  \textbf{Viscous}  & \multicolumn{2}{|c|}{\textbf{Relaxation Method}} \\
\textbf{ Stage:  }       &  \textbf{Method}   & \textbf{with $\tau_0 = 1$} &  \textbf{with $\tau_0 = 10$}  \\
\hline
before instability & $0 < t < 2.5$   & $0 < t < 3.5$   & $0 < t < 4.5$ \\
exponential growth & $2.5 < t < 4.5$ & $3.5 < t < 6.5$ & $4.5 < t < 10$ \\
saturation         & $4.5 < t < 23$  & $6.5 < t < 30 $ & $10 < t < 30 $ \\
quasi-steady state & $ t > 23$       & $ t > 30$       & $t > 30$\\
\hline
\end{tabular}
\label{table:different_phases}
\end{table}
The Kelvin-Helmholtz instability converts the kinetic energy that is available to the base horizontal shear flow into vertical fluctuations that need to overcome the stably-stratified atmosphere.
The gravitational potential energy of the system is changed during this process. In addition, after saturation the fluid starts to overturn, and is mixed, where irreversible processes change the potential energy. Here we want to investigate how the forcing contributes to the different forms of energy in the system. Using the separate components responsible for the change in different energy budgets presented above and tracking the changes of the kinetic energy, internal energy and different gravitational potential energy budgets, we will discuss how the system behaves for different forcings. Below we distinguish between four stages of the system's evolution that are: the time interval before the exponential growth of an instability, the exponential growth phase, the onset of saturation, and a fully saturated quasi-steady state. These stages are at different times for the three calculations that are discussed and can be found in Table \ref{table:different_phases}.\\

Significant differences between the forcing methods become evident when following different energy budgets normalized by the initial value of the system's total energy with time as shown in Fig. \ref{fig:figure05}. In general the sum of the three energies is increasing due to the external work done by the forcing.
%
%% $E_{kin}(t=0) = 195.81$
%
%
%viscous method in full detail
For the viscous forcing the kinetic energy remains almost constant until the instability starts to grow at $\tilde{t} \approx 3$, at that time $E_{kin}$ begins to decrease. The internal energy increases at an almost constant rate from the beginning of the calculation, which is due to viscous heating that extracts kinetic energy via $\varepsilon$ in equation (\ref{eqn:kin_energy_changes}). As the kinetic energy remains almost constant in the beginning, it can be concluded that the amount of energy dissipated is fed back into the system due to external work, $W$. The decrease in the kinetic energy of the system during the exponential growth of the instability is a direct consequence of the Kelvin-Helmholtz instability extracting energy from the large-scale shear flow in order to overcome the potential energy associated with the stably-stratified atmosphere. This amount of energy is partly converted into vertical motion, which contributes to the kinetic energy, and partly exchanged into gravitational potential energy. The term in the energy change rate associated with this exchange is $\mathcal{H}_{\rho}$, which leads to a slight increase in $E_{pot}$. Because the conversion of the mean-flow kinetic energy to vertical motion retains the energy in the kinetic energy budget, the decrease is small.\\
During the saturation phase the negative rate of change in the kinetic energy grows. Whereas for the internal energy a plateau is present just after the exponential growth phase and this is followed by a steeper increase during the end of the saturation phase. The gravitational potential energy starts to increase faster from the beginning of the saturation phase.\\
Comparing the gravitational potential energy and the internal energy shows that the changes of the internal energy are similar to the changes in the potential energy but with a time shift and a greater amplitude. The time delay is a direct consequence of the irreversible processes during the feedback of gravitational potential energy into kinetic energy, where previously kinetic energy is transformed reversibly and irreversibly into gravitational potential energy due to the term $\mathcal{H}_{\rho}$.  \\
 
%%%%%%%%%%%%%%%%%%%%%%%%%%%%%%%%%%%%%%%%%%%%%%%%%%%%%%%%%%%%%%%%%%%%%%%%%%%%%%%%%%%%%%%%%%%%%%%%%%%%%%%%%%%%%%%%%%%%%%%
%
Eventually a quasi-static state is reached, where the kinetic energy eventually will decrease very little, but the potential energies will constantly increase due to viscous heating and irreversible mixing processes. Both processes extract kinetic energy due to $\varepsilon$, $\mathcal{H}_{\rho}$, and $\mathcal{H}_{\rho}$ respectively, but only a part of $\varepsilon$ is added to the system by the external force. Therefore, the system will always evolve very slowly, but remain statistically similar for a very long time.\\
%
% \tau_0 = 10
We now focus on the calculations where the relaxation method was used to sustain a shear flow. The time evolution of $E_{kin}$ for the relaxation method with $\tau_0 = 10$ is similar to the viscous forcing. However, the early evolution is different because $E_{kin}$ decreases even before the instability starts to develop. 
This is expected since the kinetic energy initially contained in the initial shear flow is dissipated by viscosity over a short time-scale. Therefore, the exponential growth regime is shifted to later times, where a similar drop in $E_{kin}$ as was found in the viscous forcing case. In both cases, this reduction in kinetic energy corresponds to an increase in the potential energy in the system (see Fig. \ref{fig:figure05} (c)). In the non-linear regime the system also tends towards a quasi-steady state. 
Similar to the viscous forcing, for the relaxation method with $\tau_0 = 10$ the background potential energy and internal energy increase slowly until the system start to saturate. During the saturation phase a steeper increase is present. Then, after several sound crossing times, both potential energies start to converge towards a constant small growth after the system saturated. This behaviour reveals that the energy induced by the forcing principally transfers into internal energy due to dissipation, but does not contribute to an increase in either kinetic energy or available potential energy for late time evolution.\\
%
% \tau_0 = 1.0
%
A calculation with a shorter relaxation time, $\tau_0 = 1.0$, shows a different behaviour, where $E_{kin}$ increases with the onset of instability. This growth is due to the very intense external forcing present as soon as the averaged velocity profile deviates from the target profile, which is the case when the instability starts growing. For $\tau_0 = 1.0$ this corresponds to a growth in the total kinetic energy over approximately 30 sound crossing times whereby $E_{kin}$ slowly oscillates around a fixed value.\\
The case with $\tau_0 = 1.0$ shows that a constantly large increase of the potential energy is present even for the saturated stage. Looking back at Fig. \ref{fig:figure05} (a) the kinetic energy converges towards a constant value at large times. This indicates again that the kinetic energy pumped into the system by external forcing, $W$, is used to balance viscous dissipation and partly converted into gravitational potential and internal energy. Here the amount of externally added energy is significantly greater than for the other two calculations. \\
% \tau_0 = 0.1
In contrast for the calculation with $\tau_0 = 0.1$, which is not displayed here, the kinetic energy grows during the whole duration of the calculation. As discussed earlier, this growth in the kinetic energy of the system is in that particular case associated with a transition between a stably stratified atmosphere (the polytropic index is initially $m=1.6$) and a convectively unstable atmosphere where large convective cells appear in the upper part of the domain (see Fig. \ref{fig:figure04} (c)). This transition is driven by the large viscous heating in the central shear region modifying the temperature profile and changing the sign of the Brunt-V{\"a}is{\"a}l{\"a} frequency. While the interaction between a large-scale shear flow and thermal convection is of interest \citep[see for example][]{2011A&A...533A..40G, 2009MNRAS.400..337S}, this is beyond the scope of the current study. \\
%----------------------------------------potential energies---------------------------------------------------
%
%-----------------------------------------------------------------------------------------------------------
%----------------------------------------------------------------------------------------------------------------
%Available potential energy
%-------------------------------------------------------------------------------------------------------------------------------------------------------------
In Fig. \ref{fig:figure06} the evolution of the available gravitational potential energy is plotted for the three-dimensional calculations. This part of energy is due to the reversible part of $\mathcal{H}_{\rho}$ in the energy equations.
When looking at the available potential energy in Fig. \ref{fig:figure06}, no available potential energy is present before the saturation of the shear instability for all cases. Such that the system is in a state of lowest possible potential energy. The $E_{avail}$ for the viscous method and relaxation method with $\tau_0 = 10$ increases similarly, although the arch of $E_{avail}$ is shifted towards later times in the calculation with $\tau_0 = 10$. During saturation the fluid is mixed mostly, which means that due to the onset of overturning the background density is modified. This is evident from the increase of $E_{pot}$ and $E_{avail}$, which indicates reversible and irreversible mixing processes see Fig. \ref{fig:figure05} (b) and (c). After saturation less mixing occurs in the system. For the viscous method $E_{avail}$ converges towards zero for late times, whereas in the calculation using the relaxation method with $\tau_0 = 10$ the available potential energy oscillates around a small value. 
Since available potential energy is directly related to mixing \citep{doi:10.1146/annurev.fluid.35.101101.161144}, the system evolves towards a state with little mixing. This means that, after a certain modification of the density profile, the overturning settles down and persists at a low level over a long period of time.  
In agreement with kinetic energy evolution for the relaxation method with $\tau_0 = 1.0$, the available potential energy starts to growth more rapidly and the system seems to reach a type of quasi-static state at very late times. However, for both cases $\tau_0 = 1.0$ and $\tau_0 = 10$, with current limitation on numerical resources, it remains unclear if the available potential energy is saturated or will eventually decay. In order to clarify this, the calculations need to be evolved further. However, for the purposes of comparing the forcing methods in this paper it is immaterial. By using the relaxation method we can reach a long-lived state and different mixing behaviours exist depending on the relaxation time $\tau_0$, which persist sufficiently long to study long-time evolution of the generated turbulence.\\
%
%----------- work and viscous dissipation-------------------------------------------------------------------------------
%
\subsubsection{Comparing total viscous dissipation and external work}
\label{section:work_vs_visousdissipation}
\begin{figure}
  \vspace*{5pt}
\includegraphics[width=0.475\textwidth]{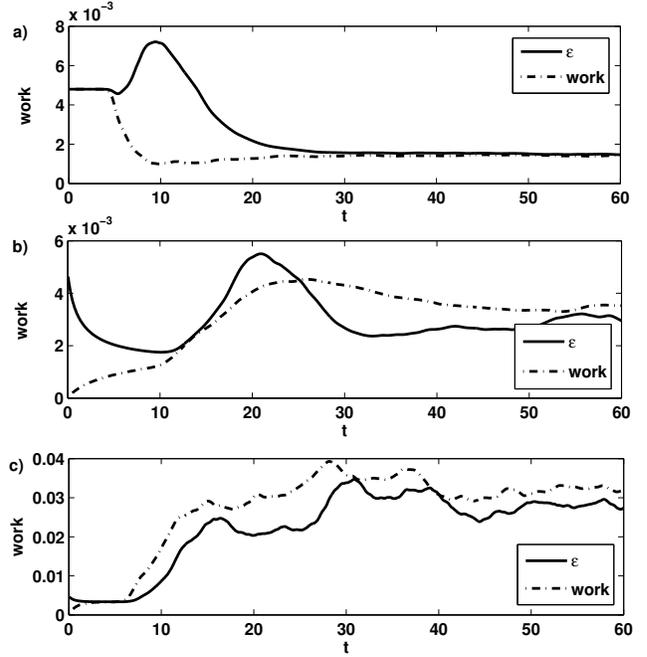}
\caption{The evolution of total viscous dissipation rate of momentum, $\varepsilon$, and the work done by the forcing, $W$ for the viscous method and the relaxation method are shown. In (a) the viscous forcing is used. The relaxation method with $\tau_0 = 10$ is used in (b), and with $\tau_0 = 1.0$ in (c).}   
\label{fig:figure07} 
\end{figure}

To investigate how much of the energy induced into the system by forcing balances the viscous dissipation, which part remains as kinetic energy and what converts into potential energy, it is useful to study the work done by the forcing, given by W as well as the total viscous dissipation rate, $\varepsilon$, with time. These quantities can be found for all three calculations in Fig. \ref{fig:figure07}. At the start of the calculation the viscous forcing will always almost exactly balance the viscous dissipation, because the velocity profile does not deviate from the target velocity such that the viscous force cancel the viscous dissipation exactly (see equation (\ref{eq:viscous_force})). This is true until approximately $\tilde{t} \approx 5$ when the instability starts to saturate. After saturation the work done by the viscous forcing is not sufficient to balance the additional dissipation associated with small-scale fluctuations in the system in that case. This is associated with a decrease in the total kinetic energy as already discussed previously. At late times the amount of $\varepsilon$  converges towards the work done (see Fig. \ref{fig:figure07} (a)) and so the system evolves towards a quasi-static state, where a sustained turbulent flow is present.\\
%--------------------------------------------------------------------------------------------------------------------------------------------------------------------
At the beginning of each calculation, using the relaxation method the work done by the forcing, $W$, is initially zero since the velocity profile exactly matches the target profile (see equation (\ref{eq:relaxation_force})). Therefore, depending on $\tau_0$, viscous dissipation is initially not balanced, as can be seen in Fig. \ref{fig:figure07} (b, c). As the initial shear flow diffuses, the associated dissipation decreases until it becomes equal to the external forcing leading to a quasi-steady state. For the case with $\tau_0 = 1.0 $ the force increases shortly after the start of the calculation such that a phase where the viscous dissipation is balanced is present before the shear flow instability start to saturate.
After saturation the work done by the forcing is greater than $\varepsilon$, which explains the increase in kinetic energy noticed previously.\\ %------------------------------------------------------------------------------------------------------------------------------------------------------------------
Due to the long relaxation time, the case with $\tau_0 = 10$ reveals a distinct behaviour, where $W$ remains less than $\varepsilon$ throughout the exponential growth phase after which it matches for a few sound-crossing times $\varepsilon$, see Fig. \ref{fig:figure07} (b). When the total viscous dissipation reaches a peak the work done remains insufficient to balance for the viscous dissipation. At large times when the system is evolving towards a quasi-static state the total viscous dissipation remains less than the energy input such that turbulence can be sustained.\\ 
%
%-------------------------------------------------------------------------------------------------------------------
%%
%
%
%
%----------------------------------------------------------------------------------------------------------------------
%Summary and discussion of the energetics
%
%----------------------------------------------------------------------------------------------------------------------
\subsubsection{Discussion}
\label{section:discussion_energetics}
Our research has revealed several characteristics of the different forcing methods: The viscous method provides a well defined localised force, but without control on the resulting velocity profile of the saturated flow. The shear instability can freely develop further away from the middle of the domain, but no turbulent motion is sustained there. Therefore, modification of the background profiles are solely due to non-linear dynamics of the instability. This results in less control on the resulting averaged velocity profile further away from the middle layer.     
From energetic considerations it can be concluded that the additional energy, that is added to the system during the late time evolution by external forcing, approaches a constant value. This initially kinetic energy is mostly converted into potential energy via dissipation. Only a small part contributes to the turbulent dynamics by mixing processes.\\
On the contrary to the viscous method using the relaxation method provides control on the resulting velocity profile, because the force is proportional to the deviation of the horizontally averaged velocity. Therefore, the target profile is controlled even far away from $z=0.5$. This corresponds to a global forcing, which can suppresses changes of the background velocity throughout the domain such that modifications of the background profile due to the shear instability are suppressed. This also can induce more mixing, which is not initially caused by the instability of the localised shear layer and therefore leads to non-physical behaviour further away from the middle plane. \\ 
An additional parameter, the relaxation time $\tau_0$, provides control on the strength of the forcing. Investigating the energy evolution revealed that decreasing $\tau_0$ results in significantly more kinetic energy induced into the system by the external force than viscous dissipation converts into internal energy. Therefore, turbulent motion is supported by the external forcing. For the choice of $\tau_0$ two typical time-scales are of interest, the turnover time $t_{s} = L_u/U_0$ and the viscous time-scale $t_{\mu} = L_u ^2 / \mu$. For $\tau_0 < t_{s}$ an instability with almost the same properties as obtained by the EV-solver is triggered (see Section \ref{sec:results_linear_compare}). However, from the energy evolution of the saturated flow it becomes obvious that only a $\tau_0$ greater than or of the same order as the viscous time-scale, which is $t_{\mu} \approx 1.6 $ for this case, leads to a system which can reach a quasi-steady state. \\
Therefore, having compared the non-linear evolution for the viscous forcing and the relaxation method, we conclude that both methods can be used depending on the properties of the dynamics that needs to be modelled. For investigations with focus on the resulting velocity profile during the saturated regime the relaxation method is more appropriate, where a careful choice of the relaxation time has to provide that no significant effects from the forcing can induce unphysical behaviour. This should be provided if $\tau_0 \sim \mathcal{O}( t_{\mu})$ or greater. On the other hand, if the non-linear evolution of a shear unstable flow is of interest, where the the mixing behaviour induced by the shear instability is the main focus, the viscous method provides a more appropriate forcing. Since the viscous method does not significantly affect the turbulent dynamics further away from the shear region, the turbulence induced by the instability can evolve freely.\\
%
%
%
%
%-------------------------------------------------------------------------------------------------------------
\section{Conclusions}
Turbulent motions driven by a shear flow instability are subject to occur in a wide range of physical systems, where numerical calculations can provide a comprehensive insight to the physical processes. Here direct numerical calculations in two- and three-dimensional Cartesian domains  are used to analyse different forcing methods, which were exploited in the past to maintain a background shear flow. In order to determine how different forcing methods affect a saturated flow and to get a more thorough understanding on possible unphysical behaviour it is essential to investigate different forcings qualitatively, e.g. by tracking global quantities and horizontally averaged profiles of velocities. \\
%------------------------------------------------------------------------------------------------------------------------------------------------------------
Testing three methods in the linear and non-linear regime reveals that two conceptually different methods, the viscous forcing and perturbation method, result in exactly the same solutions. For both of these methods the force term remains as initially set, such that there is no back reaction on the forcing by the velocity changes associated with the flow. Furthermore, in a few cases with a weak instability and moderate viscosity the unstable flow decays after saturation. The third method uses horizontally averaged velocity profiles of the actual flow to formulate a force that drives the shear flow back to the target profile after a relaxation time that can be chosen. Such a method  provides a self regulating force and more control on the strength of the forcing due to different relaxation times.\\
%--------------------------------------------------------------------------------------------------------------------------------------------------------------
Comparing the exponential growth phase with solutions from a linear stability analysis shows that the growth rates achieved by all methods used  are close to the predicted value, if for the relaxation method a sufficiently small relaxation time $\tau_0 $ is chosen. Cases with larger $\tau_0$ lead to a slightly different shear instability, since the viscous dissipation is initially not balanced and the initial state evolves before an instability can occur. However, focusing on the non-linear evolution a significant difference in the system dynamics is revealed when using the relaxation method with different $\tau_0$. When choosing $\tau_0$ greater than the viscous time-scale a non-linear evolution like for the viscous forcing method is achieved, where the long time evolution converges towards a quasi-static state. Energy induced into the system by the force balances the loss by viscous dissipation, but little additional kinetic and potential energy is obtained.
In contrary a relaxation time $\tau_0$ less than $t_{\mu}$ leads to a system which is constantly forced and develops a turbulent region which spreads across a larger region in the vertical direction. Such cases do not tend to evolve towards a quasi-static state, which becomes evident due to their energy evolution.  Moreover, the energy induced into the system is significant greater than the loss by dissipation such that the energy overrun is converted into kinetic energy of the disturbances and available potential energy of the system. \\
%--------------------------------------------------------------------------------------------------------------------------------------------------------------
Analysing the turbulent Reynolds number for late times shows that when decreasing $\tau_0$ the horizontal layer of turbulent flow reaches a larger vertical extend and very high Re numbers. However, greater $\tau_0$ and the viscous method develop a small region confined around the middle of the domain with moderate Re numbers. 
Thus the strength of the forcing has a strong impact on the spread of the resulting turbulent region. Interestingly, the mean flow resulting from viscous forcing develops a peculiar form around the middle plane, where a steep slope is present, while the relaxation method leads to a horizontally averaged velocity profile that generally preserves a hyperbolic tangent profile. Since the physical mechanism driving shear flows in different objects are not known in detail, the relaxation method provides a tool to adjust the force such that a more suitable flow can be achieved. Therefore, we conclude that the relaxation method provides a more suitable method to sustain a velocity profile when modelling stellar interior as for example the tachocline in our Sun or shear regions in more massive main-sequence stars. However, in order to study the non-linear evolution of a shear driven turbulent flow the viscous method or the equivalent perturbation method suit better, as no artificial dynamics due to the forcing affects the modification of the background profiles. \\ 
The physical mechanism for the generation and maintenance of the differential rotation in the solar interior and especially the tachocline is not well understood \citep{1998Natur.394..755G, 0004-637X-690-1-783}. It is widely believed that external processes such as Reynold stresses, which originated in the convection zone, drives the shear flow in the tachocline \citep{2008ApJ...673..557M}. However, we do not know what form the resulting force has that drives the shear flow in the tachocline. The viscous method correspond to a forcing confined within the shear region whereas the relaxation method corresponds to a bulk forcing. Prospective global-scale investigations might reveal which of these two forcings is more relevant when modelling the tachocline. \\
%--------------------------------------------------------------------------------------------------------------------------------------------------------------
Having established a detailed analysis of possible numerical methods to sustain a localised shear flow with minimised effect on the boundaries, possible applications of shear driven turbulence in stellar interiors have to be considered and are currently underway.
\section*{Acknowledgements}
This research has received funding from STFC and from the
School of Mathematics, Computer Science and Engineering at City University
London. This work partially used the ARCHER UK National Supercomputing Service (http://www.archer.ac.uk). Some of the calculations were carried out on the UK MHD Consortium computing facilities at Warwick that is supported by STFC.We are grateful to the referee for providing helpful comments on the paper.
%---------------------------------------------------------------------------------------------------------------------------------------------------------------
\bibliographystyle{mn2e}
\bibliography{bibfile}

\begin{thebibliography}{}

\bibitem[\protect\citeauthoryear{{Barker}, {Silvers}, {Proctor} \&
  {Weiss}}{{Barker} et~al.}{2012}]{2012MNRAS.424..115B}
{Barker} A.~J.,  {Silvers} L.~J.,  {Proctor} M.~R.~E.,    {Weiss} N.~O.,  2012,
  \mnras, 424, 115

\bibitem[\protect\citeauthoryear{{Brandenburg}, {Nordlund}, {Stein} \&
  {Torkelsson}}{{Brandenburg} et~al.}{1995}]{1995ApJ...446..741B}
{Brandenburg} A.,  {Nordlund} A.,  {Stein} R.~F.,    {Torkelsson} U.,  1995,
  \apj, 446, 741

\bibitem[\protect\citeauthoryear{{Br{\"u}ggen} \& {Hillebrandt}}{{Br{\"u}ggen}
  \& {Hillebrandt}}{2001}]{2001MNRAS.320...73B}
{Br{\"u}ggen} M.,  {Hillebrandt} W.,  2001, \mnras, 320, 73

\bibitem[\protect\citeauthoryear{{Brummell}, {Cline} \& {Cattaneo}}{{Brummell}
  et~al.}{2002}]{2002MNRAS.329L..73B}
{Brummell} N.,  {Cline} K.,    {Cattaneo} F.,  2002, \mnras, 329, L73

\bibitem[\protect\citeauthoryear{{Brun} \& {Toomre}}{{Brun} \&
  {Toomre}}{2002}]{2002ApJ...570..865B}
{Brun} A.~S.,  {Toomre} J.,  2002, \apj, 570, 865

\bibitem[\protect\citeauthoryear{{Caulfield} \& {Peltier}}{{Caulfield} \&
  {Peltier}}{2000}]{2000JFM...413....1C}
{Caulfield} C.~P.,  {Peltier} W.~R.,  2000, Journal of Fluid Mechanics, 413, 1

\bibitem[\protect\citeauthoryear{{Cline}, {Brummell} \& {Cattaneo}}{{Cline}
  et~al.}{2003a}]{2003ApJ...599.1449C}
{Cline} K.~S.,  {Brummell} N.~H.,    {Cattaneo} F.,  2003a, \apj, 599, 1449

\bibitem[\protect\citeauthoryear{{Cline}, {Brummell} \& {Cattaneo}}{{Cline}
  et~al.}{2003b}]{2003ApJ...588..630C}
{Cline} K.~S.,  {Brummell} N.~H.,    {Cattaneo} F.,  2003b, \apj, 588, 630

\bibitem[\protect\citeauthoryear{{Dubrulle}, {Mari{\'e}}, {Normand}, {Richard},
  {Hersant} \& {Zahn}}{{Dubrulle} et~al.}{2005}]{2005A&A...429....1D}
{Dubrulle} B.,  {Mari{\'e}} L.,  {Normand} C.,  {Richard} D.,  {Hersant} F.,
  {Zahn} J.-P.,  2005, \aap, 429, 1

\bibitem[\protect\citeauthoryear{Dudis}{Dudis}{1974}]{Dudis_1974}
Dudis J.~J.,  1974, J. Fluid Mech., 64, 65

\bibitem[\protect\citeauthoryear{Favier \& Bushby}{Favier \&
  Bushby}{2012}]{FLM:8458223}
Favier B.,  Bushby P.~J.,  2012, Journal of Fluid Mechanics, 690, 262

\bibitem[\protect\citeauthoryear{Favier \& Bushby}{Favier \&
  Bushby}{2013}]{FLM:8885996}
Favier B.,  Bushby P.~J.,  2013, Journal of Fluid Mechanics, 723, 529

\bibitem[\protect\citeauthoryear{{Favier}, {Jouve}, {Edmunds}, {Silvers} \&
  {Proctor}}{{Favier} et~al.}{2012}]{2012MNRAS.426.3349F}
{Favier} B.,  {Jouve} L.,  {Edmunds} W.,  {Silvers} L.~J.,    {Proctor}
  M.~R.~E.,  2012, \mnras, 426, 3349

\bibitem[\protect\citeauthoryear{{Goldreich} \& {Lynden-Bell}}{{Goldreich} \&
  {Lynden-Bell}}{1965}]{1965MNRAS.130..125G}
{Goldreich} P.,  {Lynden-Bell} D.,  1965, \mnras, 130, 125

\bibitem[\protect\citeauthoryear{{Gough} \& {McIntyre}}{{Gough} \&
  {McIntyre}}{1998}]{1998Natur.394..755G}
{Gough} D.~O.,  {McIntyre} M.~E.,  1998, \nat, 394, 755

\bibitem[\protect\citeauthoryear{Griffies}{Griffies}{2004}]{Griffies_climate}
Griffies S.~M.,  2004.
Princton University Press

\bibitem[\protect\citeauthoryear{{Guerrero} \& {K{\"a}pyl{\"a}}}{{Guerrero} \&
  {K{\"a}pyl{\"a}}}{2011}]{2011A&A...533A..40G}
{Guerrero} G.,  {K{\"a}pyl{\"a}} P.~J.,  2011, \aap, 533, A40

\bibitem[\protect\citeauthoryear{{Hawley}, {Balbus} \& {Winters}}{{Hawley}
  et~al.}{1999}]{1999ApJ...518..394H}
{Hawley} J.~F.,  {Balbus} S.~A.,    {Winters} W.~F.,  1999, \apj, 518, 394

\bibitem[\protect\citeauthoryear{{Heifetz}, {Mak}, {Nycander} \&
  {Umurhan}}{{Heifetz} et~al.}{2015}]{2015JFM...767..199H}
{Heifetz} E.,  {Mak} J.,  {Nycander} J.,    {Umurhan} O.~M.,  2015, Journal of
  Fluid Mechanics, 767, 199

\bibitem[\protect\citeauthoryear{{Holt}, {Koseff} \& {Ferziger}}{{Holt}
  et~al.}{1992}]{1992JFM...237..499H}
{Holt} S.~E.,  {Koseff} J.~R.,    {Ferziger} J.~H.,  1992, Journal of Fluid
  Mechanics, 237, 499

\bibitem[\protect\citeauthoryear{{Hughes} \& {Proctor}}{{Hughes} \&
  {Proctor}}{2013}]{2013JFM...717..395H}
{Hughes} D.~W.,  {Proctor} M.~R.~E.,  2013, Journal of Fluid Mechanics, 717,
  395

\bibitem[\protect\citeauthoryear{{Hughes} \& {Tobias}}{{Hughes} \&
  {Tobias}}{2001}]{2001RSPSA.457.1365H}
{Hughes} D.~W.,  {Tobias} S.~M.,  2001, Proceedings of the Royal Society of
  London Series A, 457, 1365

\bibitem[\protect\citeauthoryear{{Jacobitz}, {Sarkar} \& {van Atta}}{{Jacobitz}
  et~al.}{1997}]{1997JFM...342..231J}
{Jacobitz} F.~G.,  {Sarkar} S.,    {van Atta} C.~W.,  1997, Journal of Fluid
  Mechanics, 342, 231

\bibitem[\protect\citeauthoryear{Jones, Boronski, Brun, Glatzmaier, Gastine,
  Miesch \& Wicht}{Jones et~al.}{2011}]{Jones2011120}
Jones C.,  Boronski P.,  Brun A.,  Glatzmaier G.,  Gastine T.,  Miesch M.,
  Wicht J.,  2011, Icarus, 216, 120

\bibitem[\protect\citeauthoryear{Landau \& Lifshitz}{Landau \&
  Lifshitz}{1987}]{landau2013fluid}
Landau L.,  Lifshitz E.,  1987, Fluid Mechanics.
No.~v. 6, Pergamon Press

\bibitem[\protect\citeauthoryear{{Ligni{\`e}res}, {Califano} \&
  {Mangeney}}{{Ligni{\`e}res} et~al.}{1999}]{1999AA...349.1027L}
{Ligni{\`e}res} F.,  {Califano} F.,    {Mangeney} A.,  1999, A\& A, 349, 1027

\bibitem[\protect\citeauthoryear{Matthews, Proctor \& Weiss}{Matthews
  et~al.}{1995}]{FLM:340261}
Matthews P.~C.,  Proctor M. R.~E.,    Weiss N.~O.,  1995, J. Fluid Mech., 305,
  281

\bibitem[\protect\citeauthoryear{Miesch}{Miesch}{2003}]{0004-637X-586-1-663}
Miesch M.~S.,  2003, The Astrophysical Journal, 586, 663

\bibitem[\protect\citeauthoryear{{Miesch}, {Brun}, {De Rosa} \&
  {Toomre}}{{Miesch} et~al.}{2008}]{2008ApJ...673..557M}
{Miesch} M.~S.,  {Brun} A.~S.,  {De Rosa} M.~L.,    {Toomre} J.,  2008, \apj,
  673, 557

\bibitem[\protect\citeauthoryear{Miesch, Gilman \& Dikpati}{Miesch
  et~al.}{2007}]{0067-0049-168-2-337}
Miesch M.~S.,  Gilman P.~A.,    Dikpati M.,  2007, The Astrophysical Journal
  Supplement Series, 168, 337

\bibitem[\protect\citeauthoryear{{Narayan}, {Goldreich} \& {Goodman}}{{Narayan}
  et~al.}{1987}]{1987MNRAS.228....1N}
{Narayan} R.,  {Goldreich} P.,    {Goodman} J.,  1987, \mnras, 228, 1

\bibitem[\protect\citeauthoryear{Peltier \& Caulfield}{Peltier \&
  Caulfield}{2003}]{doi:10.1146/annurev.fluid.35.101101.161144}
Peltier W.~R.,  Caulfield C.~P.,  2003, Annual Review of Fluid Mechanics, 35,
  135

\bibitem[\protect\citeauthoryear{{Prat} \& {Ligni\`eres}}{{Prat} \&
  {Ligni\`eres}}{2013}]{refId0m01}
{Prat} V.,  {Ligni\`eres} F.,  2013, A\&A, 551, L3

\bibitem[\protect\citeauthoryear{Scinocca}{Scinocca}{1995}]{Scinocca1995}
Scinocca J.~F.,  1995, J. Atmos. Sci., 52, 2509–2530

\bibitem[\protect\citeauthoryear{Silvers}{Silvers}{2008}]{Silvers01042008}
Silvers L.~J.,  2008, Monthly Notices of the Royal Astronomical Society, 385,
  1036

\bibitem[\protect\citeauthoryear{{Silvers}, {Bushby} \& {Proctor}}{{Silvers}
  et~al.}{2009}]{2009MNRAS.400..337S}
{Silvers} L.~J.,  {Bushby} P.~J.,    {Proctor} M.~R.~E.,  2009, \mnras, 400,
  337

\bibitem[\protect\citeauthoryear{Silvers, Vasil, Brummell \& Proctor}{Silvers
  et~al.}{2009}]{1538-4357-702-1-L14}
Silvers L.~J.,  Vasil G.~M.,  Brummell N.~H.,    Proctor M. R.~E.,  2009, The
  Astrophysical Journal Letters, 702, L14

\bibitem[\protect\citeauthoryear{Smyth \& Moum}{Smyth \&
  Moum}{2000}]{/10.1063/1.870386}
Smyth W.~D.,  Moum J.~N.,  2000, Physics of Fluids, 12

\bibitem[\protect\citeauthoryear{{Smyth} \& {Winters}}{{Smyth} \&
  {Winters}}{2003}]{2003JPO....33..694S}
{Smyth} W.~D.,  {Winters} K.~B.,  2003, Journal of Physical Oceanography, 33,
  694

\bibitem[\protect\citeauthoryear{Squire}{Squire}{1933}]{Squire621}
Squire H.~B.,  1933, Proceedings of the Royal Society of London A:
  Mathematical, Physical and Engineering Sciences, 142, 621

\bibitem[\protect\citeauthoryear{Tailleux}{Tailleux}{2013}]{1402-4896-2013-T155-014033}
Tailleux R.,  2013, Physica Scripta, 2013, 014033

\bibitem[\protect\citeauthoryear{Thorpe}{Thorpe}{1987}]{JGRC:JGRC3931}
Thorpe S.~A.,  1987, Journal of Geophysical Research: Oceans, 92, 5231

\bibitem[\protect\citeauthoryear{Tobias \& Hughes}{Tobias \&
  Hughes}{2004}]{0004-637X-603-2-785}
Tobias S.~M.,  Hughes D.~W.,  2004, The Astrophysical Journal, 603, 785

\bibitem[\protect\citeauthoryear{Vasil \& Brummell}{Vasil \&
  Brummell}{2008}]{0004-637X-686-1-709}
Vasil G.~M.,  Brummell N.~H.,  2008, The Astrophysical Journal, 686, 709

\bibitem[\protect\citeauthoryear{Vasil \& Brummell}{Vasil \&
  Brummell}{2009}]{0004-637X-690-1-783}
Vasil G.~M.,  Brummell N.~H.,  2009, The Astrophysical Journal, 690, 783

\bibitem[\protect\citeauthoryear{Winters, Lombard, Riley \& D'Asaro}{Winters
  et~al.}{1995}]{FLM:354020}
Winters K.~B.,  Lombard P.~N.,  Riley J.~J.,    D'Asaro E.~A.,  1995, Journal
  of Fluid Mechanics, 289, 115

\bibitem[\protect\citeauthoryear{{Witzke}, {Silvers} \& {Favier}}{{Witzke}
  et~al.}{2015}]{2015AandAWitzke}
{Witzke} V.,  {Silvers} L.~J.,    {Favier} B.,  2015, A\&A, 577, A76

\bibitem[\protect\citeauthoryear{Zahn}{Zahn}{1974}]{Zahn_1974}
Zahn J.-P.,  1974, in Ledoux P.,  Noels A.,  Rodgers A.,  Union I.~A.,  27 I.
  A. U.~C.,   35 I. A. U.~C.,  eds, , Stellar Instability and Evolution.
Springer

\end{thebibliography}

\appendix

\label{lastpage}
\end{document}